\documentclass[journal,draftclsnofoot,onecolumn,12pt,twoside]{IEEEtranTCOM}

\normalsize

\usepackage{amsmath,amsfonts,amssymb,amsthm}
\usepackage[pdftex]{graphicx}
\usepackage{epstopdf}
\usepackage[caption=false,subrefformat=parens,labelformat=parens]{subfig}
\usepackage{multirow}
\usepackage{xcolor}
\usepackage{soul}
\definecolor{lightgray}{gray}{0.85}
\sethlcolor{lightgray}
\usepackage{algorithm}
\usepackage{algcompatible}

\algnewcommand\INPUT{\item[\textbf{Input:}]}
\algnewcommand\INITIAL{\item[\textbf{Initialization:}]}
\algnewcommand\OUTPUT{\item[\textbf{Output:}]}
\algnewcommand\RETURN{\item[\textbf{Return:}]}
\algnewcommand\ITER{\item[\textbf{Iteration:}]}
\algrenewcommand\algorithmiccomment[2][\small]{{#1\hfill\ #2}}

\begin{document}

{\Large \textbf{Notice:} This work has been submitted to the IEEE for possible publication. Copyright may be transferred without notice, after which this version may no longer be accessible.}
\clearpage

\title{Deep Neural Network Based Active User Detection for Grant-free NOMA Systems}

\author{Wonjun Kim,
        Yongjun Ahn,
        and~Byonghyo~Shim\\
        Institute of New Media and Communications and Department of Electrical and Computer Engineering, Seoul National University, Seoul, Korea
\thanks{This work was supported by 'The Cross-Ministry Giga KOREA Project' grant funded by the Korea government(MSIT) (No. GK18P0500, Development of Ultra Low-Latency Radio Access Technologies for 5G URLLC Service).}
\thanks{Parts of this paper was presented at the ICC, Shanghai, China, May 20-24, 2019~\cite{wj_ICC}.}
}


\maketitle

\vspace{-5em}
\begin{abstract}
As a means to support the access of massive machine-type communication devices, grant-free access and non-orthogonal multiple access (NOMA) have received great deal of attention in recent years. In the grant-free transmission, each device transmits information without the granting process so that the basestation needs to identify the active devices among all potential devices. This process, called an active user detection (AUD), is a challenging problem in the NOMA-based systems since it is difficult to identify active devices from the superimposed received signal. An aim of this paper is to put forth a new type of AUD based on deep neural network (DNN). By applying the training data in the properly designed DNN, the proposed AUD scheme learns the nonlinear mapping between the received NOMA signal and indices of active devices. As a result, the trained DNN can handle the whole AUD process, achieving an accurate detection of the active users. Numerical results demonstrate that the proposed AUD scheme outperforms the conventional approaches in both AUD success probability and computational complexity.
\end{abstract}

\begin{IEEEkeywords}
grant-free, non-orthogonal multiple access, active user detection, deep neural network
\end{IEEEkeywords}

\section{Introduction}
In recent years, massive machine-type communication (mMTC) has received much attention due to the variety of applications such as smart factory and building, public safety and monitoring, smart metering, to name just a few.
As the term speaks for itself, mMTC concerns the access of massive machine-type communication (MTC) devices (e.g., sensors, robots, drones, machines) to the basestation~\cite{ITU-R}.
Main goal of mMTC is to support the massive connectivity in the uplink-dominated communication.
However, this task is too demanding in the conventional wireless systems (i.e., 4G Long Term Evolution (LTE) systems) for the heavy signaling overhead caused by the complicated handshaking in the scheduling process and the lack of time/frequency resources caused by the orthogonal resource allocation to a large number of MTC devices~\cite{METIS, Scheduling}.

As a solution to support the massive connectivity, \textit{grant-free access} and \textit{non-orthogonal multiple access} have been proposed in recent years~\cite{GRANT_FREE}, \cite{LinglongDai}.
Grant-free access allows the transmission of MTC device to the basestation without the granting process.
Since each device transmits information without scheduling, identification of active devices (i.e., devices transmitting information) among all potential devices in a cell is required.
This process, often referred to as the \textit{active user detection} (AUD), is an important problem in the grant-free mMTC since without this process the basestation cannot figure out the active devices transmitting information.
In order to support the massive connectivity with limited amount of resources, an approach to use non-orthogonal sequences, called non-orthogonal multiple access (NOMA), has been proposed~\cite{LinglongDai}.
In this scheme, by the superposition of multiple devices' signals, orthogonality of transmit signals is intentionally violated.
To control the interuser interference caused by the orthogonality violation, NOMA employs device specific non-orthogonal sequences and deliberately designed nonlinear detector (e.g., successive interference cancellation (SIC) and message passing algorithm (MPA)~\cite{AMP}).

\begin{figure}[t]
	\centering
	\includegraphics[width=.75\columnwidth]{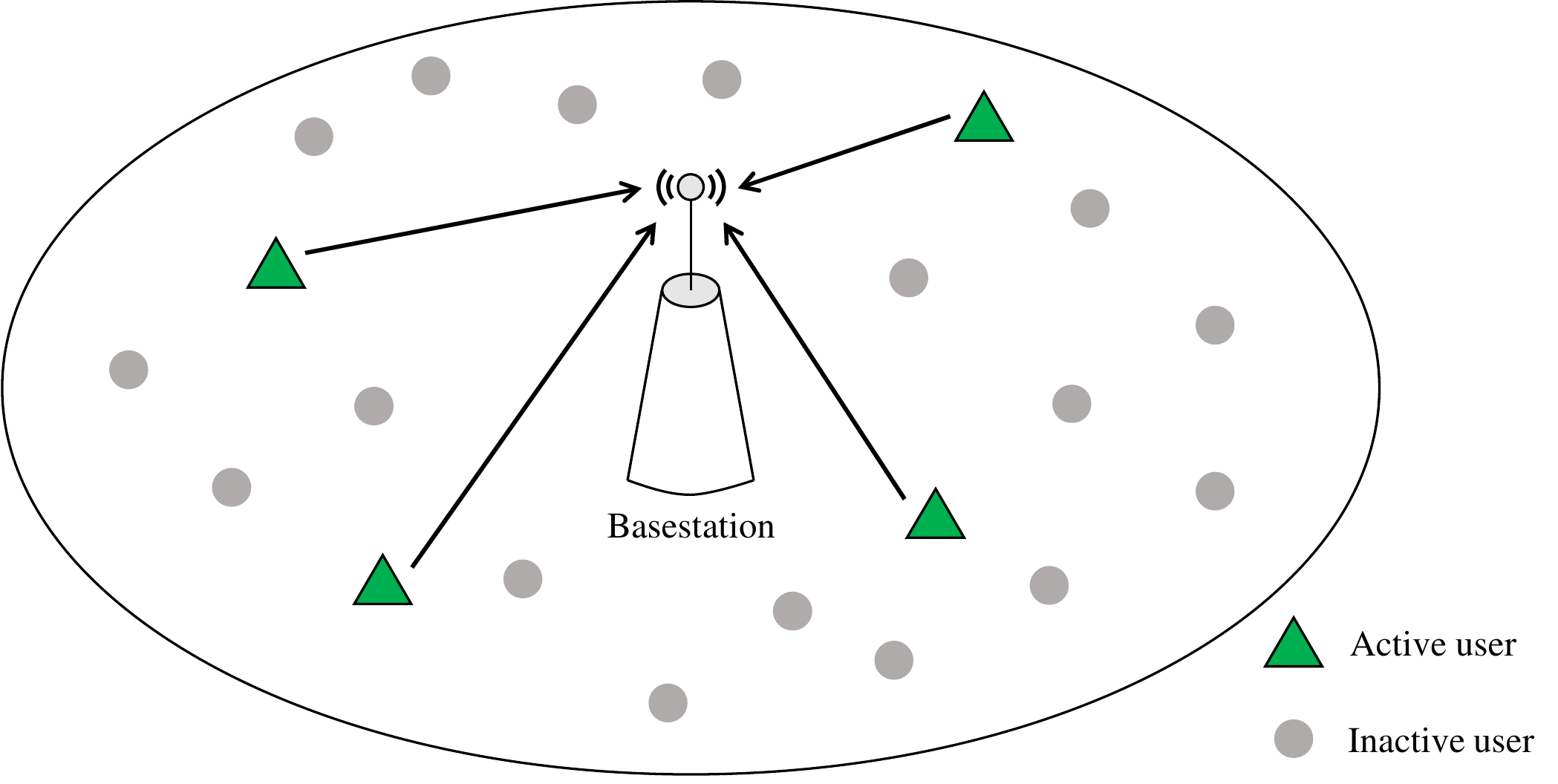}
	\caption{System model of the mMTC uplink scenario where only a few MTC devices are active.}
	\label{fig:intro}
\end{figure}

By exploiting the fact that only a few active devices in a cell transmit the information concurrently (see Fig.~\ref{fig:intro}), the AUD problem can be readily formulated as a sparse recovery problem~\cite{reviewer_1}, \cite{reviewer_2}.
Since the transmit vector is sparse, compressed sensing (CS) technique has been popularly employed \cite{Ahn,CStrick}.
In [8], the AUD problem is modeled as a single measurement vector (SMV) problem and MPA is used to solve the problem.
In this CS-based AUD scheme, basestation detects devices based on the correlation between the received signal and device specific sequence.
However, performance of the CS-based AUD is not that appealing when the columns of a system matrix (a.k.a. sensing matrix) are highly correlated and sparsity (the number of nonzero elements) of the underlying input vector increases.
In fact, in the practical NOMA-based transmission, correlation among the NOMA sequences and also device activity (sparsity) are relatively high so that the CS-based AUD might not be effective.
Indeed, it has been shown that the performance of the sparse recovery algorithm is degraded significantly when the mutual correlation and sparsity increase~\cite{CStrick}.
Therefore, it is of importance to come up with a new type of AUD scheme suitable for the overloaded yet less sparse access scenarios.

An aim of this paper is to pursue an entirely different approach to detect active users in the grant-free NOMA scenario.
For an efficient and accurate AUD, we exploit the deep neural network (DNN), a learning-based tool to approximate the complicated and nonlinear function.
Over the years, DNN has been successfully applied in numerous applications such as image classification~\cite{ImgClas_VGGNET}, machine translation~\cite{Machine_Translation}, automatic speech recognition~\cite{Speech_Recognition}, and Go game~\cite{Go_game}.
Recently, DNN has been also applied to various wireless systems such as multiple-input and multiple-output (MIMO) detection, wireless scheduling, and direction-of-arrival (DoA) estimation~\cite{Deep_MIMO_Detection}.
In these works, DNN is used to learn a desired nonlinear function (e.g., classification and decision) through the training process.
In [16], for instance, the DNN network learns the mapping between the interference pattern and the optimized scheduling.
In our framework, DNN learns the complicated mapping between the received NOMA signal and the indices of active users in the transmit signal.
To be specific, the proposed AUD scheme, henceforth referred to as deep AUD (D-AUD), learns the sparse structure of device activity from a set of training data.
It is now well-known from the \textit{universal approximation theorem} that DNN processed by the deeply stacked hidden layers can well approximate the desired function~\cite{Univ_Approx_Theorem}.
In our context, this means that the trained DNN with multiple hidden layers can handle the whole AUD process, resulting in an accurate detection of the active users.

From the numerical evaluations on the grant-free NOMA systems, we demonstrate that the proposed D-AUD scheme outperforms the conventional CS-based approaches by a large margin in terms of the AUD success probability and computational complexity.
In particular, in realistic mMTC environments specified in 3rd Generation Partnership Project (3GPP) New Radio (NR) Rel. 16~\cite{3GPPNOMA}, we observe that the D-AUD scheme achieves more than 4 dB gain in terms of the AUD success probability (see Fig. 9) and 80\% reduction in computational complexity over the conventional approaches (see Table 1).
Note that the complexity reduction is achieved by the fact that operations in the test phase are just simple addition and multiplication.

The main contributions of this paper are as follows:
\begin{itemize}
\item We propose a novel AUD scheme based on the deep neural network and explain the detailed operation of D-AUD network components. To be specific, we train the D-AUD scheme to learn the mapping between the received NOMA signal and user activity pattern. By feeding the massive training data into the properly designed DNN, we can design an accurate and robust AUD system.
\item We tackle various issues for the D-AUD implementation. Particularly, in order to reduce the overhead in the training data collection, we propose an offline training strategy using the synthetically generated data. We also propose a sparsity estimation technique using the trained internal parameters.
\item We provide a complexity analysis and empirical simulation results from which we demonstrate the computational gain of the D-AUD over the conventional approaches.
\end{itemize}

The rest of this paper is organized as follows.
In Section II, we describe the system model and conventional AUD scheme.
In Section III, we discuss the proposed D-AUD scheme and provide a detailed description of the neural network.
Various implementation issues are discussed in Section IV.
In Section V, we present simulation results to verify the performance gain of the proposed technique and conclude the paper in Section VI.

We briefly summarize notations used in this paper. We employ uppercase boldface letters for matrices and lowercase boldface letters for vectors.
The operations $(\cdot)^{T}$ and $(\cdot)^{H}$ denote the transpose and conjugate transpose, respectively.
The operators $\odot$ and $\oslash$ denote the Hadamard product and the Hadamard division, respectively.
$\mathbb{C}$ and $\mathbb{R}$ denote the field of complex numbers and real numbers, respectively.
Also, $\mathbb{N}$ denotes the field of natural numbers.
$\lVert \cdot \rVert_{p}$ indicates the $p$-norm.
$\langle \mathbf{a}, \mathbf{b} \rangle$ is an inner product between $\mathbf{a}$ and $\mathbf{b}$.
$\Re(c)$ and $\Im(c)$ are the real and imaginary part of $c$, respectively.
$\mathbf{x}_{i}$ denotes the $i$-th column of the matrix $\mathbf{X}$ and $x_i$ is the $i$-th element of the vector $\mathbf{x}$.
$\mathbf{X}_{\Omega}$ is the submatrix of $\mathbf{X}$ that contains the columns specified in the set $\Omega$ and $\mathbf{x}_{\Omega}$ is the vector constructed by picking the elements specified in the set $\Omega$.
$\mathbf{A}^{\dagger}$ is the Penrose-Moore inverse of the matrix $\mathbf{A}$.

\section{AUD System Model}
\begin{figure}[t]
	\centering
	\includegraphics[width=.95\columnwidth]{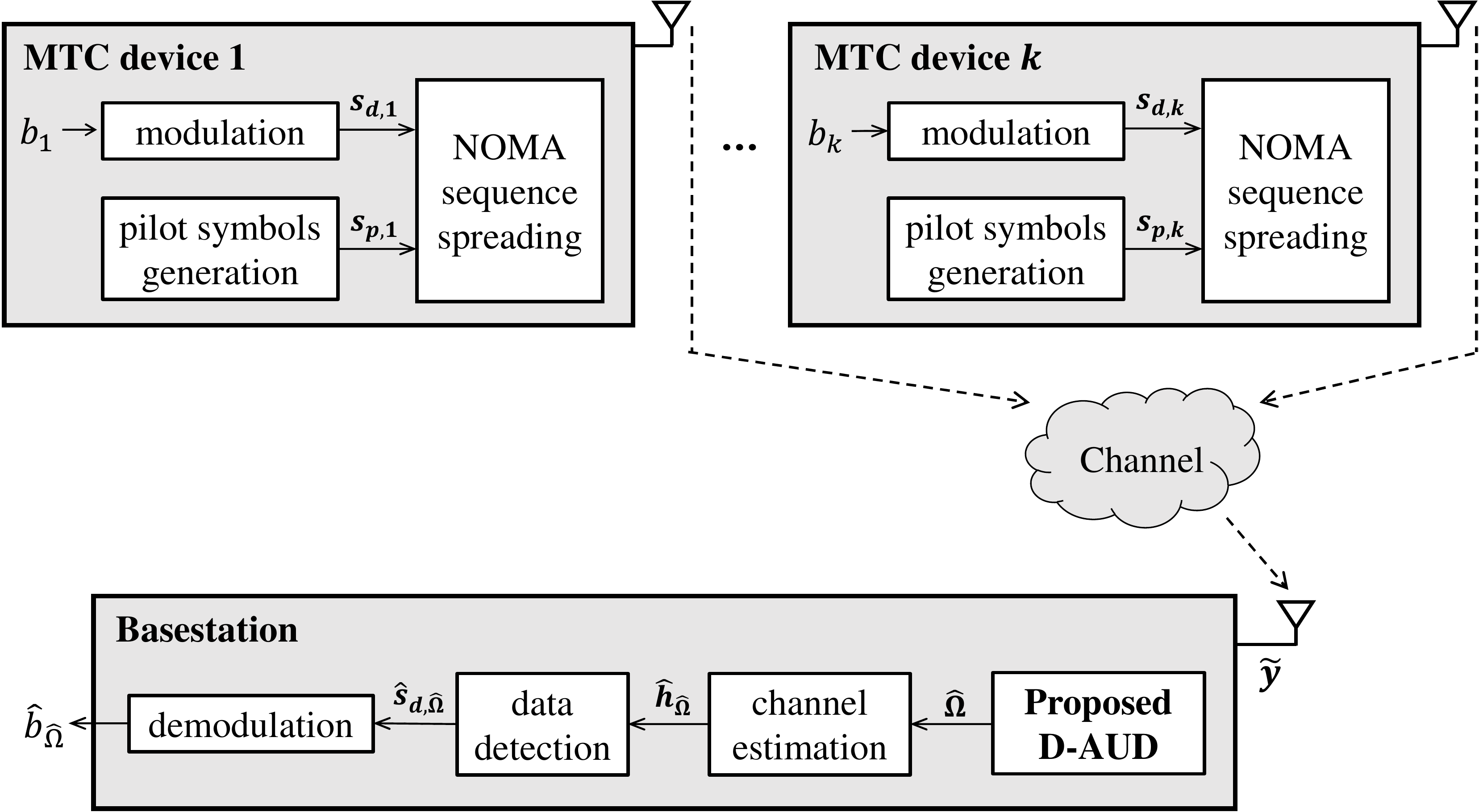}
	\caption{Block diagram of the proposed D-AUD scheme.}
	\label{fig:SystemModel}
\end{figure}

We consider the uplink grant-free NOMA systems in which the basestation equipped with a single antenna receives information from multiple machine-type devices with a single antenna\footnote{Extension of the system model to the multi-antenna model is straightforward (see Section V. B).}
In particular, we consider the overloaded scenario where the number of devices $N$ in a cell is larger than the number of frequency resources $m$ ($m<N$).
Since each device can transmit packets freely without scheduling, the basestation should identify \textit{active} devices transmitting packets.
Active devices transmit both pilot and data symbols after the spreading with the device specific (non-orthogonal) sequences\footnote{ Basically, there are two types of sequence selection approaches in grant-free NOMA: 1) random sequence selection and 2) preconfigured sequence selection. In the random sequence selection, collision event can occur since users select a sequence randomly. Whereas, in the preconfigured sequence selection, the basestation assigns sequence to the mobile device via the random access procedure so that the collisions caused by the duplicated sequences can be prevented. In this work, we use the preconfigured sequence selection.} (see Fig. \ref{fig:SystemModel}).
Specifically, the bitstream is mapped to the symbol $s_{i}$ and then converted to the spreading vector $\mathbf{q}_{i}=\mathbf{c}_{i}s_{i}$ using the device specific codeword $\mathbf{c}_{i}$.

In this work, we employ the low-density signature (LDS) sequence where the codeword of a device contains lots of zeros~\cite{Hoshyar}.
Due to the sparse nature of a codeword, each symbol is spread into only a small number of resources, resulting in the reduction of the interuser interference.
For example, the LDS codebook $\mathbf{C}_{(4,6)}$ when $6$ devices transmit information using $4$ resources is
\begin{eqnarray}
\mathbf{C}_{(4,6)} =
\left[\begin{matrix}
0 & w_0 & w_1 & 0 & w_2 & 0 \\
w_0 & 0 & w_2 & 0 & 0 & w_1  \\
0 & w_1 & 0 & w_2 & 0 & w_0 \\
w_2 & 0 & 0 & w_1 & w_0 & 0
\end{matrix}\right]\label{lds},
\end{eqnarray}
where $w_{j}$ is the non-zero element of the codeword~\cite{Hoshyar}.

Let $s_{i}$ be the transmit symbol for the $i$-th device, then the observation vector $\mathbf{y}$ at the basestation is given by
\begin{align}
\mathbf{y}
&=
\sum_{i=1}^{N}\operatorname{diag}(\mathbf{c}_{i})\mathbf{h}_{i}s_{i}+\mathbf{v} \\
&=
\mathbf{C}\mathbf{q}+\mathbf{v},
\label{eq:single_measurement}
\end{align}
where $\mathbf{c}_{i}=[c_{i,1} \ \cdots \ c_{i,m}]^{T}$ is the LDS codeword vector for the $i$-th device, $\mathbf{h}_{i}=[h_{i,1} \ \cdots \ h_{i,m}]^{T}$ is the channel vector between the $i$-th device and the basestation, $\mathbf{v}\sim\mathcal{CN}(0,\sigma^2\mathbf{I})$ is the complex Gaussian noise vector, $\mathbf{C}=\left[ \operatorname{diag}(\mathbf{c}_{1}) \ \cdots \ \operatorname{diag}(\mathbf{c}_{N}) \right]$ is the codebook matrix of all devices in a cell, and $\mathbf{q}=\left[\mathbf{q}_{1}^{T} \ \cdots \ \mathbf{q}_{N}^{T} \right]^{T}=\left[ (s_{1}\mathbf{h}_{1})^{T} \  \cdots \ (s_{N}\mathbf{h}_{N})^{T} \right]^{T}$ is the composite of symbol and channel vectors.
It is worth pointing out that $\mathbf{q}_{i}$ contains the (frequency-domain) channel vector $\mathbf{h}_{i}$.
Since only a few devices are active at a given time, the vector $\mathbf{q}$ can be readily modeled as a sparse vector.

In performing the AUD, we use multiple, say $N_{d}$, data measurements.
Let $\tilde{\mathbf{y}} = \big[\left(\mathbf{y}^{(1)}\right)^{T} \ \cdots \ $ 
$\left(\mathbf{y}^{(N_{d})}\right)^{T} \big]^{T}$ be the stacked vector of the $N_{d}$ measurements, then
\begin{align}
\tilde{\mathbf{y}}
&=
\begin{bmatrix}
\mathbf{C}^{(1)} & \mathbf{0} & \cdots & \mathbf{0}  \\
\mathbf{0} & \mathbf{C}^{(2)} & \cdots & \mathbf{0} \\
\vdots & \vdots & \ddots &\vdots \\
\mathbf{0} & \mathbf{0} & \cdots & \mathbf{C}^{(N_{d})}
\end{bmatrix}
\begin{bmatrix}
\mathbf{q}^{(1)} \\ \mathbf{q}^{(2)} \\ \vdots \\ \mathbf{q}^{(N_{d})}
\end{bmatrix}
+
\begin{bmatrix}
\mathbf{v}^{(1)} \\ \mathbf{v}^{(2)} \\ \vdots \\ \mathbf{v}^{(N_{d})}
\end{bmatrix},
\end{align}
where $\mathbf{C}^{(t)}=\left[\operatorname{diag}(\mathbf{c}_{1}^{(t)}) \ \cdots \  \operatorname{diag}(\mathbf{c}_{N}^{(t)}) \right]$ and $\mathbf{q}^{(t)}=\left[ (s_{1}^{(t)}\mathbf{h}_{1}^{(t)})^{T} \  \cdots \ (s_{N}^{(t)}\mathbf{h}_{N}^{(t)})^{T} \right]^{T}$ is the vector whose element is the composite of the channel vector and data symbol.
Since the indices of active devices are the same for all $\mathbf{q}^{(t)}$, the supports\footnote{If $\mathbf{s} = [0 \ 1 \ 0 \ 0 \ 1 \ 0]$, then the support of $\mathbf{s}$ is $supp(\mathbf{s})=\{2,5\}$.} of $\mathbf{q}^{(t)}$ for $t=1,\ldots,N_{d}$ will also be the same (i.e., $supp(\mathbf{q}^{(1)})=supp(\mathbf{q}^{(2)})=\cdots$).
In order to identify the active device, therefore, it would be better to re-arrange the system model based on the index of devices.
To this end, we use a device activity indicator $\delta_{i}$ where $\delta_{i}=1$ for the active device and $\delta_{i}=0$ for the rest (inactive device).
Using the device activity indicator, the received vector $\tilde{\mathbf{y}}$ can be expressed as
\begin{align}
\tilde{\mathbf{y}}
=
\left[\begin{matrix}\mathbf{\Phi}_{1}& \ \cdots \ &\mathbf{\Phi}_{N}\end{matrix}\right]\left[\begin{matrix}\delta_{1}\mathbf{x}_{1} \\ \vdots \ \\ \delta_{N}\mathbf{x}_{N}\end{matrix}\right]+\left[\begin{matrix}\mathbf{v}^{(1)} \\ \ \vdots \ \\ \mathbf{v}^{(N_{d})}\end{matrix}\right]
=
\mathbf{\Phi}\mathbf{x}+\mathbf{v},
\label{eq:system_model}
\end{align}
 where ${\mathbf{x}}_{i}=\left[{(s_{i}^{(1)}\mathbf{h}_{i}^{(1)}})^{T} \ \cdots \ {(s_{i}^{(N_{d})}\mathbf{h}_{i}^{(N_{d})}})^{T}\right]^{T}$ and $\mathbf{\Phi}_{i}=\left[\operatorname{diag}( \mathbf{c}_{i}^{(1)}) \ \cdots \ \operatorname{diag}( \mathbf{c}_{i}^{(N_{d})})\right]$ are the re-arranged sparse vector and codebook matrix for the $i$-th device, respectively, $\mathbf{\Phi} = \left[\begin{matrix}\mathbf{\Phi}_{1} & \ \cdots \ &\mathbf{\Phi}_{N}\end{matrix}\right]$, and $\mathbf{x}=[\delta_{1}\mathbf{x}_{1}^{T} \ \cdots \ \delta_{N}\mathbf{x}_{N}^{T}]^{T}$.

Since a small number of devices (say $k$ devices) are active, the stacked sparse vector $\mathbf{x}$ has $k$ nonzero blocks, which implies that the received vector $\tilde{\mathbf{y}}=\mathbf{\Phi}\mathbf{x}+\mathbf{v}$ can be expressed as a linear combination of $k$ submatrices of $\mathbf{\Phi}_{1},\cdots,\mathbf{\Phi}_{N}$ perturbed by the noise.
Note that $\mathbf{\Phi}$ is available at the basestation since all entries of the codebook matrix $\mathbf{C}$ are known in advance.
In light of this, main task of the basestation is to identify the submatrices $\mathbf{\Phi}_{i}$ in $\mathbf{\Phi}$ participating in $\tilde{\mathbf{y}}$.
For example, if the second and fifth devices are active (i.e., $\Omega=\{2,5\}$), then $\mathbf{\Phi}_{2}$ and $\mathbf{\Phi}_{5}$ participate in $\tilde{\mathbf{y}}$.
Note that this setup is standard in the compressed sensing~\cite{CStrick}.
The corresponding AUD problem can be formulated as the support identification problem as
\begin{align}
\tilde{\Omega} = \arg\min_{\vert \Omega \vert = k}\frac{1}{2}\left\Vert \tilde{\mathbf{y}} - \mathbf{\Phi}_{\Omega}\mathbf{x}_{\Omega}  \right\Vert_{2}^{2}.
\label{eq:AUD_model}
\end{align}
In solving \eqref{eq:AUD_model}, greedy sparse recovery algorithm such as block orthogonal matching pursuit (BOMP)~\cite{CStrick} and block compressive sampling matching pursuit (B-CoSaMP)~\cite{BCOSAMP} can be used.
In each iteration, greedy sparse recovery algorithm identifies one submatrix of $\mathbf{\Phi}$ at a time using a greedy strategy.
In $j$-th iteration, for example, a submatrix $\mathbf{\Phi}_{l}$ of $\mathbf{\Phi}$ that is maximally correlated with the residual vector $\mathbf{r}^{j-1}$ is chosen.
An index of the nonzero submatrix of $\mathbf{\Phi}$ chosen at $j$-th iteration is
\begin{align}
\omega_{j}=\arg\max_{l=1,\cdots,N} \left\Vert \mathbf{\Phi}^{H}_{l}\mathbf{r}^{j-1} \right\Vert_{2}^{2},
\label{eq:bomp_iteration}
\end{align}
where $\mathbf{r}^{j-1}=\mathbf{y}-\mathbf{\Phi}_{\Omega^{j-1}}\hat{\mathbf{x}}^{j-1}$ is the $j$-th residual vector and $\hat{\mathbf{x}}^{j-1}=\mathbf{\Phi}_{\Omega^{j-1}}^{\dagger}\mathbf{y}$ is the estimate of $\mathbf{x}$ at $(j-1)$-th iteration.
One can easily see that the support identification performance depends heavily on the correlation between the residual $\mathbf{r}^{(\cdot)}$ and sensing matrix $\mathbf{\Phi}$ generated from the codebook matrix $\mathbf{C}$.

After identifying the support $\Omega$, a basestation detects the symbol vector $\hat{\mathbf{s}}_{\hat{\Omega}}$ of the active device.
To be specific, by removing the components associated with the non-support elements in \eqref{eq:system_model}, the system model can be converted from the underdetermined system  to the overdetermined system $(m>k)$.
For example, if the identified support is $\Omega=\{2,5\}$, then the system model in \eqref{eq:system_model} can be simplified to $\tilde{\mathbf{y}} = \begin{array}{cc} [\mathbf{\Phi}_{2} & \mathbf{\Phi}_5] \end{array} \left[\begin{array}{c} \mathbf{x}_{2} \\ \mathbf{x}_{5} \end{array}\right] + \mathbf{v}$ and thus a conventional technique such as the linear minimum mean square error (LMMSE) estimator followed by the symbol slicer can be used for the symbol detection (see Fig.~\ref{fig:SystemModel}).

In real scenarios, this type of CS-based AUD schemes might not be effective for the following reasons.
First, correlation of codewords increases with the number of devices.
Indeed, when we try to support a large number of devices using small amount of resources, column dimension of the codebook $\mathbf{C}$ would be much larger than the size of measurement vector $\tilde{\mathbf{y}}$, increasing the underdetermined ratio $\frac{N}{m}$ of the system.
In this case, clearly, the mutual coherence\footnote{The mutual coherence $\mu\left(\mathbf{\Phi}\right)$ is defined as the largest magnitude of normalized inner product between two distinct columns of $\mathbf{\Phi}$~\cite{CStrick}: $$ \mu\left(\mathbf{\Phi}\right) = \max_{i \neq j}\frac{\left\vert \langle \mathbf{\Phi}_{i}, \mathbf{\Phi}_{j} \rangle \right\vert}{\lVert \mathbf{\Phi}_{i} \rVert_{2} \lVert \mathbf{\Phi}_{j} \rVert_{2} }. $$} of $\mathbf{C}$ will increase sharply, causing a severe degradation of the AUD performance.
Second, when the activity of devices is high (i.e., $k$ is large), required number of iterations of the greedy sparse recovery algorithm will also increase.
Recalling that the residual vector is updated using the estimated support in each iteration (see \eqref{eq:bomp_iteration}), an error caused by the incorrectly chosen support element will be propagated (this phenomenon is called \textit{error propagation}), deteriorating the AUD performance severely.
Last but not least, computational complexity and latency of the iterative algorithm are burdensome in the real-time AUD since the complexity and processing time of sparse recovery algorithm depend heavily on the number of active devices\footnote{For example, the computational complexity of BOMP is in the order of $m^{2}kN$. Therefore, increase in the number of active devices will directly affect the computational complexity.}.
Due to the reasons mentioned, when the number of active devices is large, the CS-based AUD scheme would not be an appealing solution.
Without doubt, design of new type of AUD scheme robust to the codeword correlation and high device activity is of great importance for the success of grant-free NOMA systems in 5G and beyond\footnote{Various NOMA proposals (e.g., power-domain NOMA, LDS-OFDM, and SCMA) have been proposed in 3GPP Rel. 15~\cite{3GPPNOMA} and standardization effort is still underway.}.

\begin{figure*}[t]
	\centering
    \includegraphics[width=\linewidth]{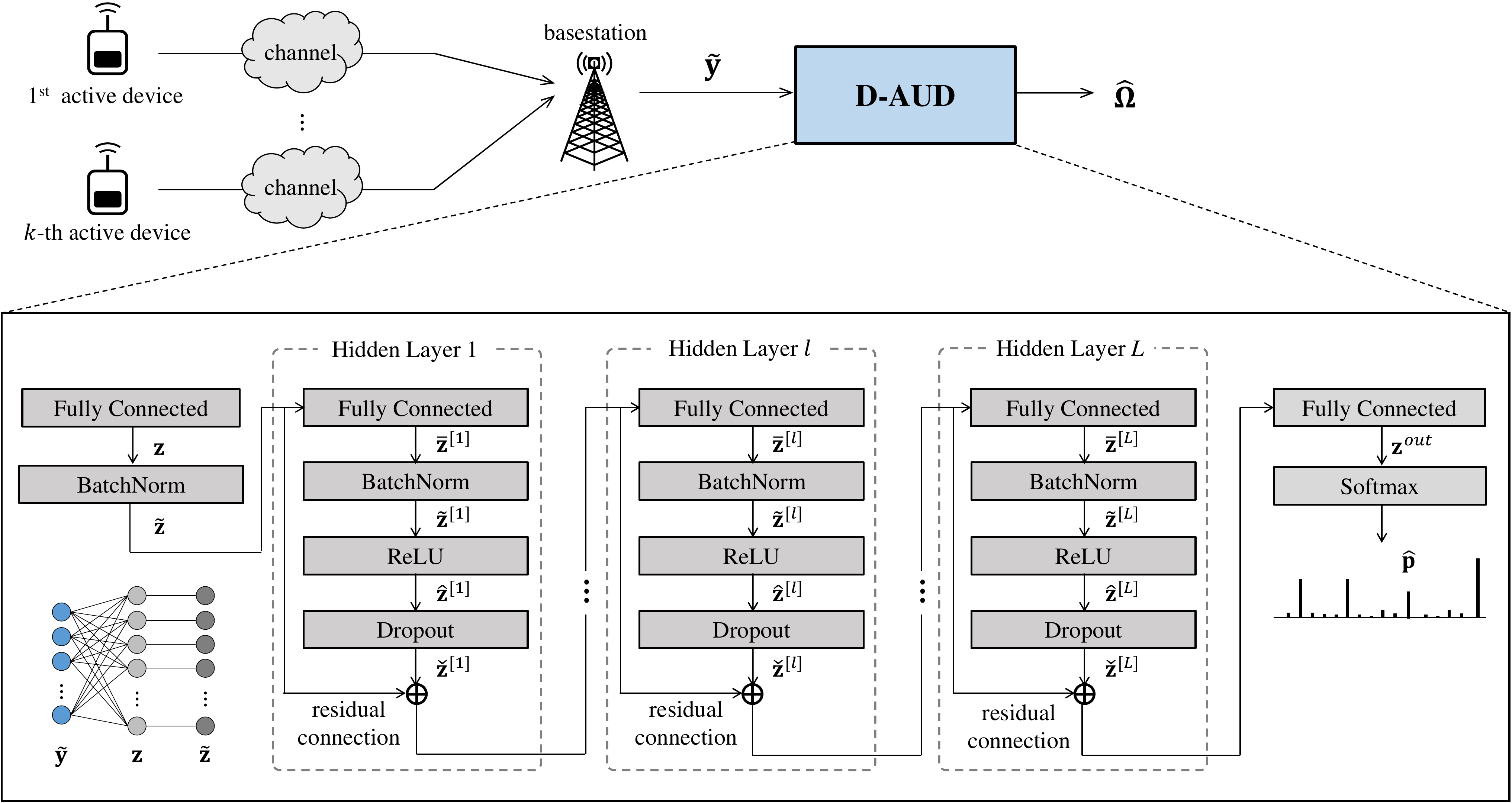}
\caption{Detailed architecture of the proposed D-AUD.}
\label{fig:DNN_AUD}
\end{figure*}

\section{Deep Neural Network Based AUD}
As mentioned, main goal of AUD is to identify the nonzero positions of $\mathbf{x}$, not the recovery of nonzero elements.
In this work, we use DNN to achieve the goal.
DNN is a feedforward neural network having multiple hidden units between input and output~\cite{review_5}.
By feeding the training data into the properly designed DNN and using the backpropagation process, we learn the nonlinear mapping $g$ between the input (i.e., received signal vector $\tilde{\mathbf{y}}$) and the support of $\mathbf{x}$.
The resulting support identification problem of the proposed D-AUD can be expressed as
\begin{align}
\hat{\Omega} = g(\tilde{\mathbf{y}}; \Theta),
\end{align}
where $\tilde{\mathbf{y}}$ is the input vector and $\Theta$ is the set of weights and biases of D-AUD network.

\subsection{D-AUD Architecture}
The primary task of the D-AUD is to find out $g$ parameterized by $\Theta$ given $\tilde{\mathbf{y}}$, closest to the optimal mapping function $g^{\ast}$.
Fig.~\ref{fig:DNN_AUD} depicts the structure of the proposed D-AUD technique.
D-AUD consists of multiple building blocks including fully-connected (FC) layers, rectified linear unit (ReLU) layer, dropout layer, and softmax layer with the batch normalization. 
In the training process, we use $P$ training data $\tilde{\mathbf{y}}^{(1)}, \cdots, \tilde{\mathbf{y}}^{(P)}$ in each training iteration.
Since $\tilde{\mathbf{y}}^{(p)}$ is a complex vector, we split the real and imaginary parts and use $\hat{\mathbf{y}}^{(p)}=[\Re{(\hat{y}_{1}^{(p)})} \ \cdots \ \Re{(\hat{y}_{m}^{(p)})} \ \Im{(\hat{y}_{1}^{(p)})} \ \cdots \ \Im{(\hat{y}_{m}^{(p)})}]$ as an input vector.
The output vector $\mathbf{z}^{(p)} \in \mathbb{R}^{\alpha\times 1}$ of the FC layer can be expressed as\footnote{$\alpha$ is a hyper-parameter representing the width of hidden layers. In general, when $\alpha$ is large, the training performance becomes high due to the large learning capacity. We will more discuss $\alpha$ in Section V.}
\begin{align}
    \mathbf{z}^{(p)} = \mathbf{W}^{in}\hat{\mathbf{y}}^{(p)}+\mathbf{b}^{in}, \qquad \text{for } p=1,\cdots,P,
\label{eq:initial}
\end{align}
where $\mathbf{W}^{in} \in \mathbb{R}^{\alpha \times 2m}$ is the initial weight and $\mathbf{b}^{in} \in \mathbb{R}^{\alpha \times 1}$ is the initial bias.
After the FC layer, $P$ output vectors are stacked in the mini-batch $\mathbf{B}=\left[ \mathbf{z}^{(1)} \ \cdots \ \mathbf{z}^{(P)} \right]^{T}$ and then normalized.
This process is referred to as the batch normalization~\cite{BN}.
In this step, each element $z_{i}^{(p)}$ ($i=1,\cdots,\alpha$) in $\mathbf{B}$ is normalized to have zero mean and unit variance.
Then, the normalized element is scaled and shifted by internal parameters.
The output $\tilde{\mathbf{z}}^{(p)}$ of the batch normalization is expressed as
\begin{align}
    \tilde{z}_{i}^{(p)} = \beta \left(\frac{z_{i}^{(p)}-\mathbf{\mu}_{\mathbf{B},i}}{\sqrt{\mathbf{\sigma}^{2}_{\mathbf{B},i}}}\right) + \gamma, \qquad \text{for } i=1,\cdots,\alpha,
    \label{eq:batch_normalization}
\end{align}
\noindent
where $\mathbf{\mu}_{\mathbf{B},i}=\frac{1}{P}\sum_{p=1}^{P}z_{i}^{(p)}$ and $\mathbf{\sigma}^{2}_{\mathbf{B},i}=\frac{1}{P}\sum_{p=1}^{P}(z_{i}^{(p)}-\mu_{\mathbf{B},i})^{2}$ are the batch-wise mean and variance, respectively, $\beta$ is the scaling parameter, and $\gamma$ is the shifting parameter.
One can see that this normalization process enforces the input distribution to have the fixed means and variances.
When the variation of input data is large, it is difficult to extract internal features (e.g., block sparse structure and codebook structure) from the input data.
Indeed, since mobile devices in different wireless geometries transmit the data in grant-free NOMA scenario, variation in $\tilde{\mathbf{y}}$ is typically very large.
Using the batch normalization, therefore, D-AUD can handle the variation of inputs caused by the different channel state and noise level.

\begin{figure}[t]
	\centering
    \includegraphics[width=.6\linewidth]{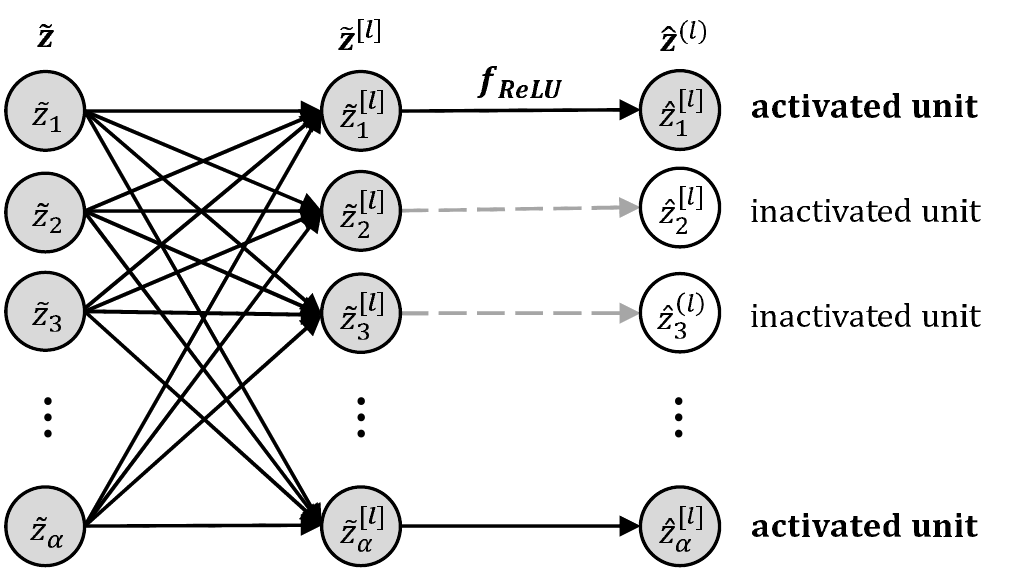}
\caption{Description of the ReLU layer.}
\label{fig:relu}
\end{figure}

After the batch normalization, the output vector $\tilde{\mathbf{z}}$ passes through the multiple hidden layers\footnote{In the sequel, we omit the training data index $p$ for notational simplicity.}.
Each hidden layer consists of the FC layer, batch normalization layer, ReLU layer, dropout layer with a residual connection\footnote{The key feature of residual connection is to put the direct identity (shortcut) connection between the stacked hidden layers. To be specific, denoting the input $\mathbf{x}$ and the desired underlying mapping as $H(\mathbf{x})$, the multiple hidden layers are fit to the residual mapping $F(\mathbf{x})=H(\mathbf{x})-\mathbf{x}$, not $H(\mathbf{x})$ directly. Since the input vector is directly linked to the output of hidden layer, the information (feature) can be delivered across the hidden layers without distortion and attenuation. Hence, we can achieve a reduction in the training error.} (see Fig.~\ref{fig:DNN_AUD}).
The output of the $l$-th FC layer $\bar{\mathbf{z}}^{[l]}$ is given by
\begin{align}
    \bar{\mathbf{z}}^{[l]} =  \mathbf{W}^{[l]}\left(\tilde{\mathbf{z}} + \sum_{i=1}^{l-1}\check{\mathbf{z}}^{[i]}\right)+\mathbf{b}^{[l]},
\label{eq:iterative_FC}
\end{align}
where $\mathbf{W}^{[l]} \in \mathbb{R}^{\alpha \times \alpha}$ and $\mathbf{b}^{[l]} \in \mathbb{R}^{\alpha \times 1}$ are the weight and bias in the $l$-th FC layer, respectively and $\check{\mathbf{z}}^{[i]}=f\left( \beta^{[i]} \left( \mathbf{W}^{[i]} \left(  \tilde{\mathbf{z}} + \sum_{j=1}^{i-1} \check{\mathbf{z}}^{[j]} \right)+\mathbf{b}^{[i]}-\boldsymbol{\mu}^{[i]}\right)\oslash \boldsymbol{\sigma}^{[i]} + \gamma^{[i]}\right)\odot \mathbf{d}^{[i]}$ is the output of the $i$-th dropout layer (we will say more about this in the next page)\footnote{$\boldsymbol{\mu}^{[i]}=[\frac{1}{P}\sum_{p=1}^{P}\bar{z}_{1}^{(p)},\cdots,\frac{1}{P}\sum_{p=1}^{P}\bar{z}_{\alpha}^{(p)}]^{T}$ and $\boldsymbol{\sigma}^{[i]}=\left[\sqrt{\frac{1}{P}\sum_{p=1}^{P}(\bar{z}_{1}^{(p)}-\mu^{[i]}_{1})^{2}}, \cdots, \sqrt{\frac{1}{P}\sum_{p=1}^{P}(\bar{z}_{\alpha}^{(p)}-\mu^{[i]}_{\alpha})^{2}}\right]^{T}$.}.
Then, the batch normalization is performed to reduce the variation of $\bar{\mathbf{z}}^{[l]}$.
After that, a nonlinear activation function is applied to $\tilde{\mathbf{z}}^{[l]}$ to determine whether the information $(\tilde{z}^{[l]}_{1},\cdots,\tilde{z}^{[l]}_{\alpha})$ generated by the hidden unit is activated (delivered to the next layer) or not (see Fig.~\ref{fig:relu})~\cite{ReLU}.
To this end, an activation function such as the sigmoid function or ReLU function can be used and thus
\begin{align}
    \hat{\mathbf{z}}^{[l]} = f(\tilde{\mathbf{z}}^{[l]}).
\end{align}

\begin{figure}[t]
\centering
\subfloat[]{\includegraphics[width=.3\linewidth]{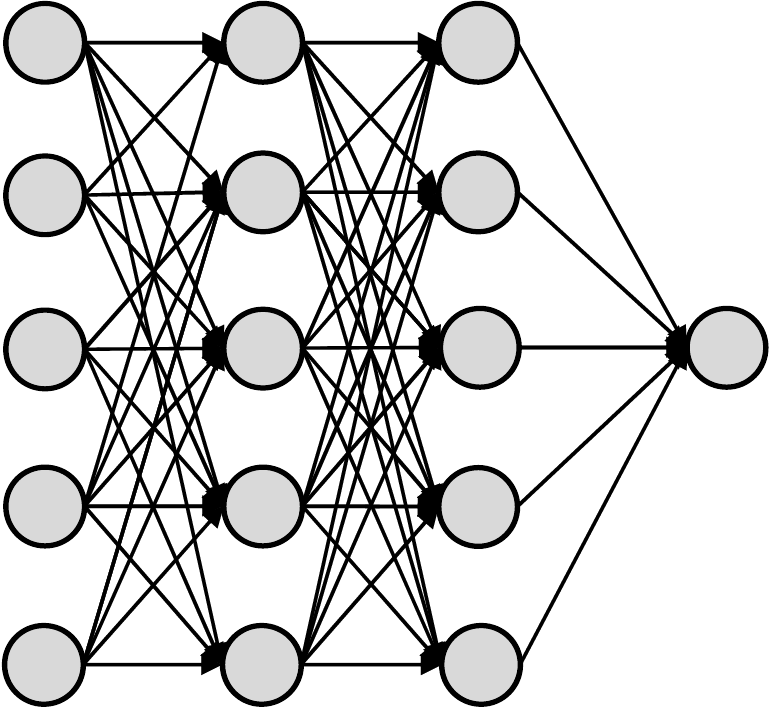}}
\hspace{3em}
\subfloat[]{\includegraphics[width=.3\linewidth]{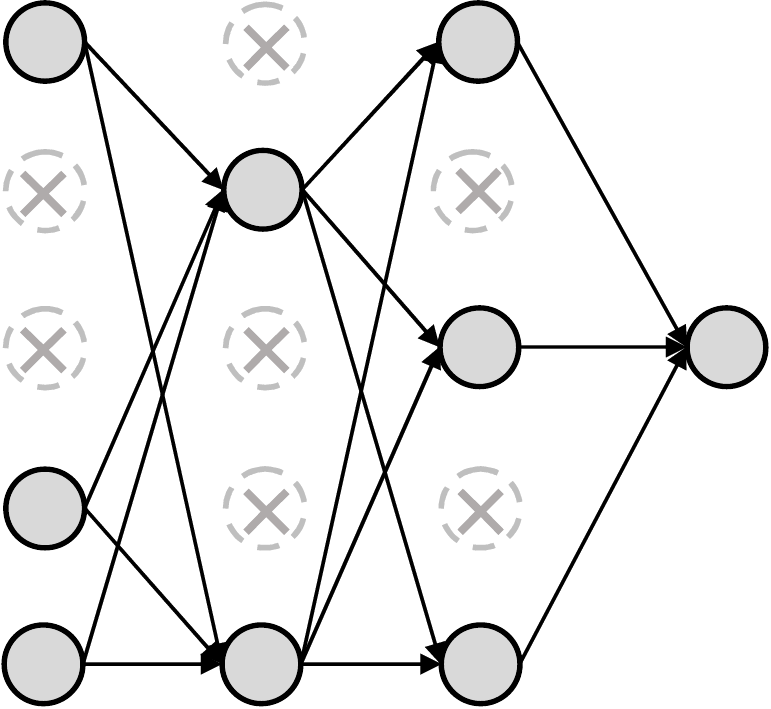}}
\caption{Dropout neural network model: (a) A standard neural network consists of three hidden layers. All hidden units in hidden layers are activated. (b) After applying the dropout, the activated hidden units are dropped out randomly.}
\label{fig:dropout}
\end{figure}

\begin{figure}[t]
	\centering
    \includegraphics[width=\linewidth]{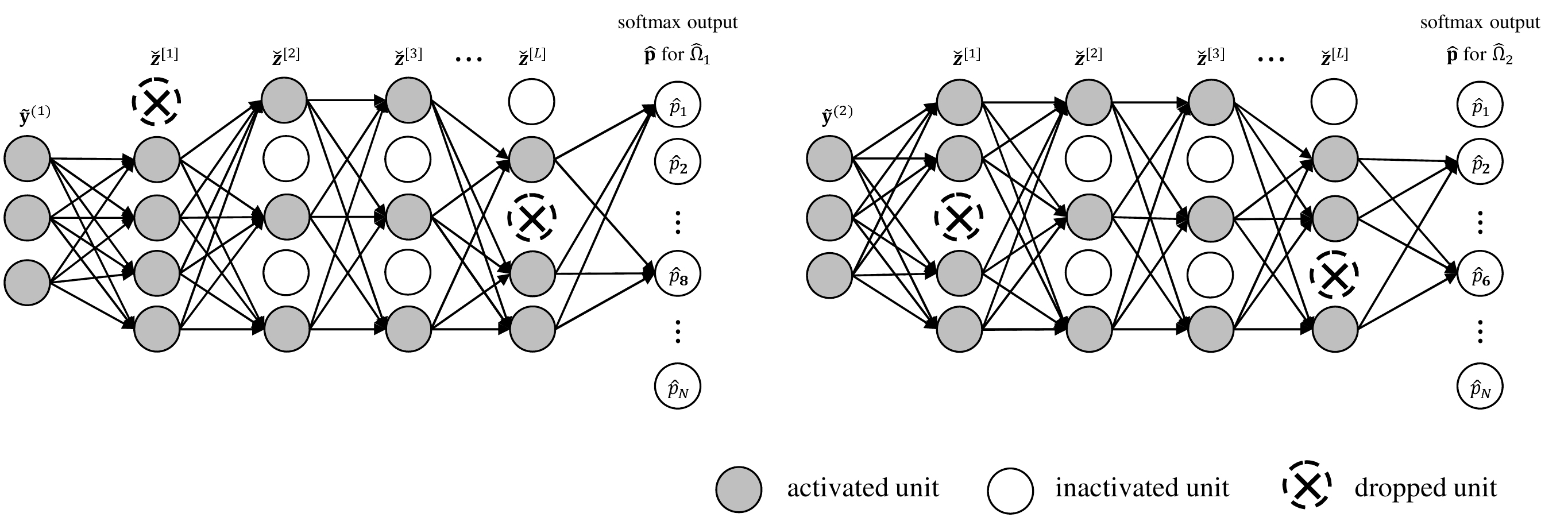}
\caption{Examples of activation patterns corresponding to the strongly correlated supports $\Omega_{1}$ and $\Omega_{2}$. Using the dropout layer, the randomly chosen hidden units are dropped out and the activation patterns for $\Omega_{1}$ and $\Omega_{2}$ can be better resolved.}
\label{fig:ex_dropout}
\end{figure}

Since the proposed scheme learns the mapping between $\tilde{\mathbf{y}}$ and the support $\Omega$, an estimate of the support $\hat{\Omega}$ would be strongly affected by the activation patterns, presumably on/off patterns, of hidden units.
When the sensing matrix $\mathbf{\Phi}$ is less correlated (i.e., the mutual coherence of the sensing matrix $\mathbf{\Phi}$ is low), $\tilde{\mathbf{y}}$ can be expressed as a linear combination of \textit{less correlated} columns of $\mathbf{\Phi}_{\Omega}$, and thus the identification of $\Omega$ from $\tilde{\mathbf{y}}$ would be relatively easy and straightforward.
Whereas, when the sensing matrix $\mathbf{\Phi}$ is highly correlated, mapping between $\tilde{\mathbf{y}}$ and $\Omega$ might not be clear and can be easily confused in the presence of randomly distributed perturbations (e.g., channel estimation error, inter-user interference, and noise).
Suppose two columns of $\mathbf{\Phi}$ are strongly correlated and only one of these is associated with the support, then it might not be easy to distinguish a correct support element from an incorrect one.
For example, if $\Omega_{1}=\{1,8\}$ and $\Omega_{2}=\{2,6\}$ and $\left\vert \langle \mathbf{\Phi}_{1},\mathbf{\Phi}_{2} \rangle \right\vert \approx 1$ and $\left\vert \langle \mathbf{\Phi}_{8},\mathbf{\Phi}_{6} \rangle \right\vert \approx 1$, then the activation patterns of hidden units for $\Omega_{1}$ and $\Omega_{2}$ would be quite similar, ending up having incorrect support identification even in the presence of a small perturbation.
In order to reduce this type of mistake, we use the \textit{dropout} layer where the activated hidden units are dropped out randomly (see Fig.~\ref{fig:dropout})~\cite{Dropout}.
By removing incoming and outgoing connections of the dropped units, similarity (ambiguity) of the activation patterns among correlated supports can be better resolved, which implies that D-AUD can identify the support accurately (see the illustration in Fig.~\ref{fig:ex_dropout}).

Let $\mathbf{d}^{[l]}$ be the dropout vector, then the $i$-th element $d_{i}^{[l]}$ of $\mathbf{d}^{[l]}$ and the final output of the $l$-th hidden layer are 
\begin{align}
    d_{i}^{[l]} &\sim Bern(\mathrm{P}_{drop}) \\
    \check{\mathbf{z}}^{[l]} &= \mathbf{d}^{[l]} \odot \hat{\mathbf{z}}^{[l]}
\end{align}
where $Bern(\mathrm{P}_{drop})$ is a Bernoulli random variable which takes the value 0 with the dropout probability $\mathrm{P}_{drop}$ and 1 with the probability $1-\mathrm{P}_{drop}$.
For example, if the second and fifth hidden units are dropped out, then $\mathbf{d}^{[l]}=[1 \ 0 \ 1 \ 1 \ 0 \ 1 \ \cdots \ 1]$ and hence $\check{z}_{2}^{[l]}$ and $\check{z}_{5}^{[l]}$ are 0.

After passing through the $L$ hidden layers, the output FC layer produces $N$ output values whose dimension is matched with the number of total users.
The output vector $\mathbf{z}^{out}$ is given by
\begin{align}
\mathbf{z}^{out} = \mathbf{W}^{out} \left(\tilde{\mathbf{z}} + \sum_{i=1}^{L}\check{\mathbf{z}}^{[i]}\right) + \mathbf{b}^{out},
\label{eq:output_FC}
\end{align}
where $\mathbf{W}^{out} \in \mathbb{R}^{N \times \alpha}$ and $\mathbf{b}^{out} \in \mathbb{R}^{N \times 1}$ are the corresponding weight and bias, respectively.
Then, the softmax layer maps $N$ output values into $N$ probabilities ($\hat{p}_{1},\cdots,\hat{p}_{N}$) representing the likelihood of being the true support element.
The $i$-th probability $\hat{p}_{i}$ is given by
\begin{align}
\hat{p}_{i} = \frac{e^{z^{out}_{i}}}{\sum\limits_{j=1}^{N} e^{z^{out}_{j}}}, \qquad \text{for } i=1,\cdots,N.
\label{eq:softmax_output}
\end{align}
Finally, an estimate of the support $\hat{\Omega}$ is obtained by taking $k$ elements having the largest probabilities:
\begin{align}
    \hat{\Omega} = \arg\max_{\left\vert \Omega \right\vert=k}{\sum_{i \in \Omega} \hat{p}_{i}}.
    \label{eq:final_support}
\end{align}

\subsection{D-AUD Training}
In the training phase, we use the training set to find out the network parameter set $\Theta^{*}$ minimizing the loss function $J(\Theta)$ (i.e., $\Theta^{*} = \arg\min_{\Theta} J(\Theta)$).
When the loss function $J(\Theta)$ is differentiable, network parameters can be updated by the gradient descent method in each training iteration.
Specifically, parameters in the $j$-th training iteration $\Theta_{j}$ are updated simultaneously in the direction of the steepest descent as
\begin{align}
   \Theta_{j}=\Theta_{j-1}-\eta\nabla_{\Theta}J(\Theta),
\end{align}
where $\nabla_{\Theta}J(\Theta)$ is the gradient of $J(\Theta)$ with respect to $\Theta$ and $\eta$ is the learning rate determining the step size.

Recalling that the final output of the D-AUD scheme is the $N$-dimensional vector $\hat{\mathbf{p}}$ whose element represents the probability of being the support element, $\hat{\mathbf{p}}=[\hat{p}_{1}, \cdots, \hat{p}_{N}]$ needs to be compared against the true probability $\mathbf{p}$ in the loss function calculation.
Since $k$ active users are assumed to be equiprobable, we set the true probability as $p_{i}=\frac{1}{k}$ for $i\in\Omega$ and $p_{i}=0$ for the rest.
For example, when the second and fourth devices are active (i.e. $k=2$ and $\Omega=\{2,4\}$), $p_{2}=p_{4}=\frac{1}{2}$ and $p_{i}=0, i\notin\{2,4\}$.
In the generation of the loss function, we use the cross entropy loss $J(\mathbf{p}, \hat{\mathbf{p}})$ defined as
\footnote{
Kullback-Leibler (KL) divergence can be also used for the loss function.
Using the KL divergence loss $D_{KL}(\mathbf{p}\parallel\hat{\mathbf{p}})$, we have
\begin{align}
D_{KL}(\mathbf{p}\parallel\hat{\mathbf{p}})
= \sum_{i=1}^{N}p_{i}\log\frac{p_{i}}{\hat{p}_{i}}
= -\sum_{i=1}^{N}p_{i}\log\hat{p}_{i}+\sum_{i=1}^{N}p_{i}\log p_{i}
= J(\mathbf{p}, \hat{\mathbf{p}}) + \sum_{j=1}^{k}p_{\omega_{j}}\log p_{\omega_{j}}
= J(\mathbf{p}, \hat{\mathbf{p}}) - \log k.
\end{align}
Since $\log k$ is a constant, to minimize the cross entropy loss $J(\mathbf{p}, \hat{\mathbf{p}})$ is essentially the same as to minimize the KL divergence loss. The regression loss function (e.g., mean square error (MSE) and the mean absolute error (MAE)), however, might not be suitable for the D-AUD training since it is used for estimating a specific value.}
\begin{align}
J(\mathbf{p}, \hat{\mathbf{p}})
=
-\sum_{i=1}^{N} p_{i}\log\hat{p}_{i}
=-\frac{1}{k}\sum_{j=1}^{k} \log\hat{p}_{\omega_{j}},
\label{eq:cost_func}
\end{align}
where $\omega_{j}\in\Omega$.
In order to minimize \eqref{eq:cost_func}, $\sum_{j=1}^{k} \log\hat{p}_{\omega_{j}}$ should be maximized.
Since the sum of softmax output values is 1 (i.e., $\sum_{i} \hat{p}_{i}=1$), the maximum can be achieved when $\hat{p}_{\omega_{j}}=\frac{1}{k}$, which is the desired training result.

\begin{figure}[t]
\centering
\subfloat[]{\includegraphics[width=.3\linewidth]{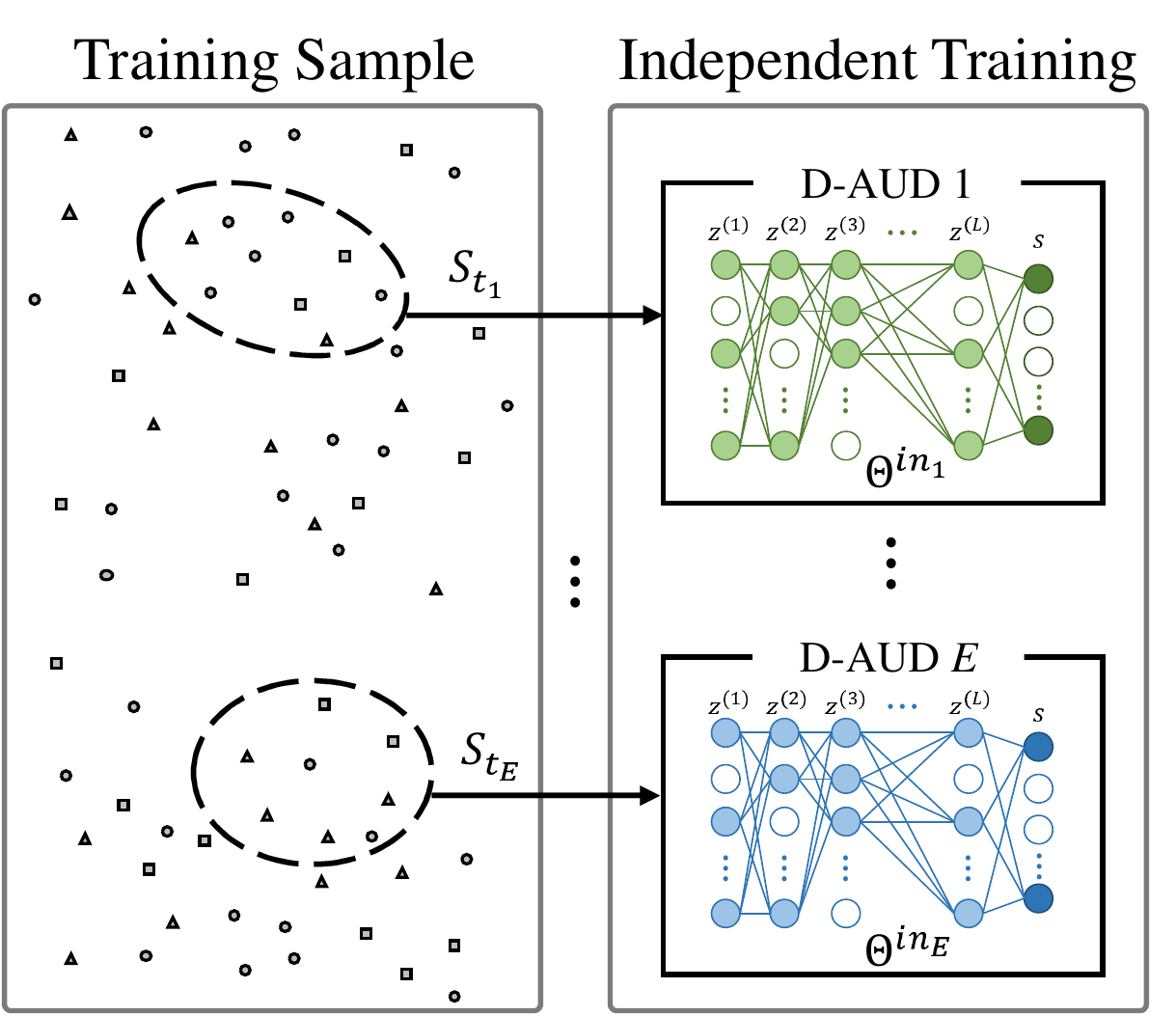}}
\subfloat[]{\raisebox{0.6cm}{\includegraphics[width=.5\linewidth]{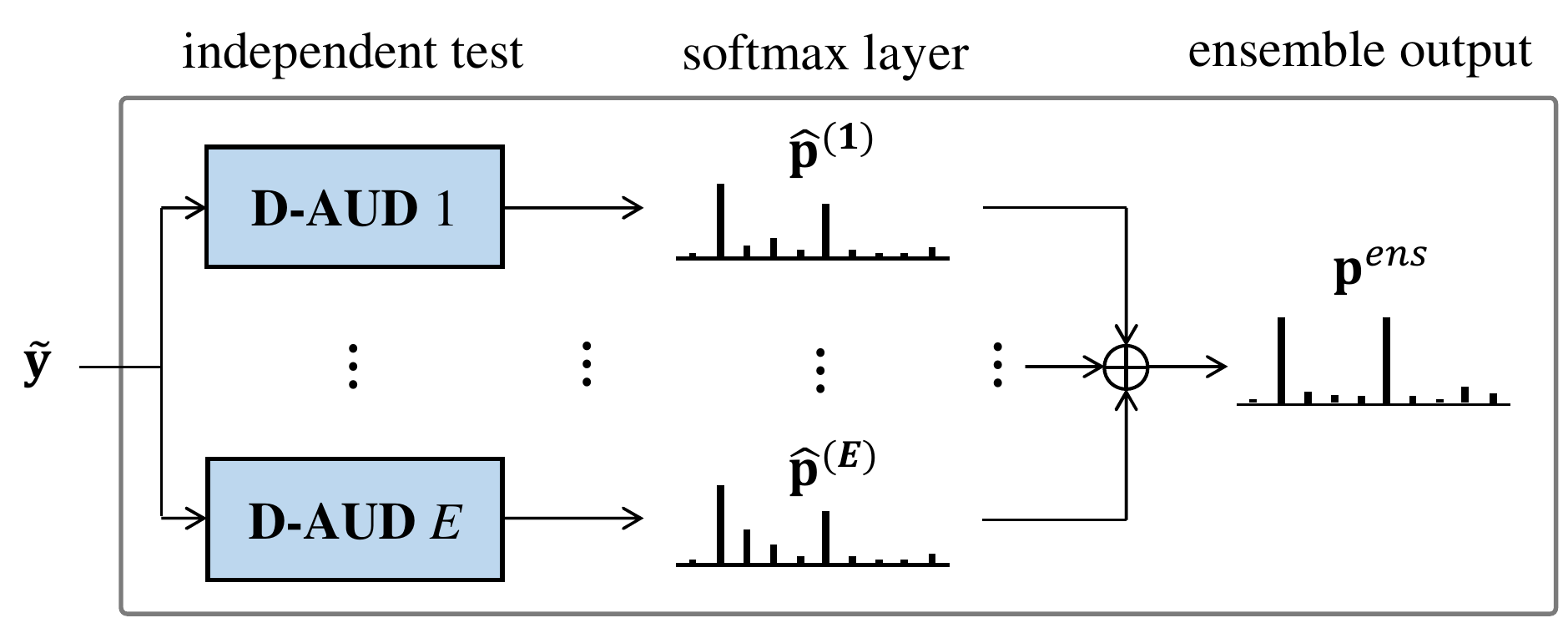}}}
\caption{Description of the ensemble network: (a) training phase for independent D-AUD scheme with different training set and (b) emsembling test phase using the independently trained D-AUD schemes.}
\label{fig:ensemble}
\end{figure}

One potential problem occurring in the training of DNN is the overfitting.
By overfitting, we mean that the designed D-AUD is so closely fitted to the training set and thus it does not make reasonable prediction for newly observed data.
Indeed, when a user not participated in the training process transmits a packet, the overfitted neural network might fail to detect this user.
In order to address this issue, we use multiple independently trained networks in the output generation.
In this scheme, called \textit{ensemble} technique~\cite{ensemble}, multiple, say $E$, D-AUD networks are trained independently with the different training sets ($S_{t_{1}}, \cdots, S_{t_{E}}$) and initial parameters ($\Theta^{in_{1}}, \cdots, \Theta^{in_{E}}$) (see Fig.~\ref{fig:ensemble}).
Thus, from the same set of measurements, $E$ independent output probabilities ($\hat{\mathbf{p}}^{(1)}, \cdots, \hat{\mathbf{p}}^{(E)}$) are generated.
By averaging out these probabilities, we obtain the ensemble probability $\mathbf{p}^{ens}$ as
\begin{align}
\mathbf{p}^{ens} = \frac{1}{E}\sum_{j=1}^{E}\hat{\mathbf{p}}^{(j)}.
\end{align}
Finally, an estimate of the support is obtained by picking indices of $k$ largest values in $\mathbf{p}^{ens}$.
One can observe that the ensemble technique is conceptually analogous to the receiver diversity technique in wireless communication systems in the sense that it is done in basestation side and also does not require additional wireless resources (e.g., frequency, time, and transmission energy) in the mobile side.

\subsection{Comments on Complexity}
In this subsection, we analyze the computational complexity of the proposed D-AUD scheme.
In our analysis, we measure the complexity in terms of the number of floating point operations (flops).
Initially, in the FC layer, the input vector $\hat{\mathbf{y}} \in \mathbb{R}^{2m\times 1}$ is multiplied by the initial weight $\mathbf{W}^{in}\in \mathbb{R}^{\alpha \times 2m}$ and the bias $\mathbf{b}^{in}\in \mathbb{R}^{\alpha\times 1}$ is added (see~\eqref{eq:initial}).
The complexity of the initial FC layer $\mathcal{C}_{\mathrm{in}}$ is
\begin{align}
\mathcal{C}_{\mathrm{in}} = (4m-1)\alpha + \alpha = 4m\alpha.
\label{eq:com_initial}
\end{align}
Since the element-wise scalar multiplication and addition are performed twice in the batch normalization process (see~\eqref{eq:batch_normalization}), the complexity $\mathcal{C}_{\mathrm{BN}}$ of batch normalization is simply
\begin{align}
\mathcal{C}_{\mathrm{BN}} = 4\alpha.
\end{align}

Next, in the hidden layer, an input vector is multiplied by the weight $\mathbf{W}^{[l]}\in \mathbb{R}^{\alpha \times \alpha}$ and then the bias $\mathbf{b}^{[l]}\in \mathbb{R}^{\alpha\times 1}$ is added (see~\eqref{eq:iterative_FC}).
After the batch normalization ($4\alpha$ flops), we test whether the value is larger than 0 or not for each element using the ReLU function.
The dropout vector $\mathbf{d}^{[l]}$ is multiplied to $\hat{\mathbf{z}}^{[l]}$ (see (14)) and then an output vector of the previous hidden layer is added to the output of the dropout layer for the residual connection.
Therefore, the complexity $\mathcal{C}_{\mathrm{hide}}$ of $L$ hidden layers can be expressed as
\begin{align}
\mathcal{C}_{\mathrm{hide}} = L\left\{(2\alpha-1)\alpha + \alpha + 4\alpha+\alpha+\alpha+\alpha\right\} = 2L\alpha^2 + 7L\alpha.
\end{align}
After passing through $L$ hidden layers, the weight multiplication and bias addition are performed in the output FC layer (see~\eqref{eq:output_FC}).
Since $\mathbf{W}^{out}\in \mathbb{R}^{N \times \alpha}$ and $\mathbf{b}^{out}\in \mathbb{R}^{N}$, the complexity $\mathcal{C}_{\mathrm{out}}$ of the output FC layer is
\begin{align}
\mathcal{C}_{\mathrm{out}} = (2\alpha-1)N + N = 2\alpha N.
\end{align}
Next, the softmax operation consisting of exponential computation ($N$ flops), summation ($N-1$ flops), and division ($N$ flops) is performed (see~\eqref{eq:softmax_output}).
The resulting computational complexity of the softmax operation is
\begin{align}
\mathcal{C}_{\mathrm{softmax}} = 3N-1.
\label{eq:com_softmax}
\end{align}
Finally, the complexity $\mathcal{C}_{\mathrm{sort}}$ of taking $k$ largest probabilities in $\mathbf{p}$ (see in~\eqref{eq:final_support}) is~\cite{GOMP}
\begin{align}
\mathcal{C}_{\mathrm{sort}} = kN - \frac{k(k+1)}{2}.
\label{eq:com_sorting}
\end{align}
From~\eqref{eq:com_initial} to~\eqref{eq:com_sorting}, the complexity $\mathcal{C}_{\mathrm{D-AUD}}$ of D-AUD is summarized as
\begin{align}
\mathcal{C}_{\mathrm{D-AUD}}
&=
\mathcal{C}_{\mathrm{in}} + \mathcal{C}_{\mathrm{BN}} + \mathcal{C}_{\mathrm{hide}} + \mathcal{C}_{\mathrm{out}} + \mathcal{C}_{\mathrm{softmax}} + \mathcal{C}_{\mathrm{sort}} \\
&=
2L\alpha^2 + (4m+7L+2N+4)\alpha + (k+3)N-\frac{k(k+1)}{2}-1.
\end{align}


\begin{table}[]
\centering
\caption{Comparison of Computational Complexity $(N=80, m=40, \alpha=500, L=6)$}
\begin{tabular}{|c||c|c|c|c|}
\hline
\multirow{2}{*}{} & \multirow{2}{*}{the number of floating point operations (flops)} & \multicolumn{3}{c|}{Complexity for various sparsity} \\ \cline{3-5} 
                  &                                                    & k=6      & k=8       & k=10      \\ \hline
\textbf{D-AUD}          &     $2L\alpha^2+ (4m+7L+2N+4)\alpha + (k+3)N-\frac{k(k+1)}{2}-1$                                               & $4.99\times 10^{6}$       & $5.59\times 10^{6}$       & $6.19\times 10^{6}$       \\
& $+ 2m + k\left( \frac{14}{3}m^{3}+m^{2}-m \right)$ & & & \\ \hline
\textbf{MMSE-BOMP}              &$2km^2 N-k + 2km + \frac{k(k+1)}{2} \left( \frac{14}{3}m^3+3m^2-m \right) + k(k+1)m^2$                                      & $7.91\times 10^{6}$         & $1.30\times 10^{7}$       & $1.92\times 10^{7}$       \\ \hline
\textbf{LS-BOMP}          &      $2km^2 N + \frac{k^4+6k^3+7k^2+2k}{12} m^3 + k(k+1)m^2 -k$                                              & $1.68\times 10^{7}$       & $4.29\times 10^{7}$       & $9.19\times 10^{7}$ \\ \hline
\end{tabular}
\end{table}

In Table I, we compare the complexities of D-AUD, MMSE-BOMP, and LS-BOMP (see Appendix A for the detailed complexity derivation).
For fair comparison, in the D-AUD, we add the complexity of the MMSE estimation $\mathcal{C}_{MMSE}=2m+k\left(\frac{14}{3}m^{3}+m^{2}-m\right)$ for the signal detection.
In order to examine the overall behavior, we compute the required flops for various sparsity levels ($k=6,8,10$).
We observe that the complexity of D-AUD is much smaller than that of conventional approaches.
For example, when $k=8$, the complexity of the D-AUD is 57\% and 87\% lower than those of MMSE-based BOMP and LS-based BOMP, respectively.
It is worth mentioning that the complexity of D-AUD depends heavily on the DNN network parameters ($L$ and $\alpha$), not the system parameters ($k$ and $N$).
For instance, when $k$ increases from $6$ to $10$, the computational complexity of D-AUD changes slightly but that of LS-BOMP increases significantly.
Thus, in the practical NOMA-based environment where the numbers of total users and active users (e.g., $N=100$ and $k=10$) are large, the D-AUD scheme can achieve a significant reduction in complexity.

\section{Practical Issues for D-AUD Implementation}
\begin{figure}[t]
\centering
\includegraphics[width=.6\columnwidth]{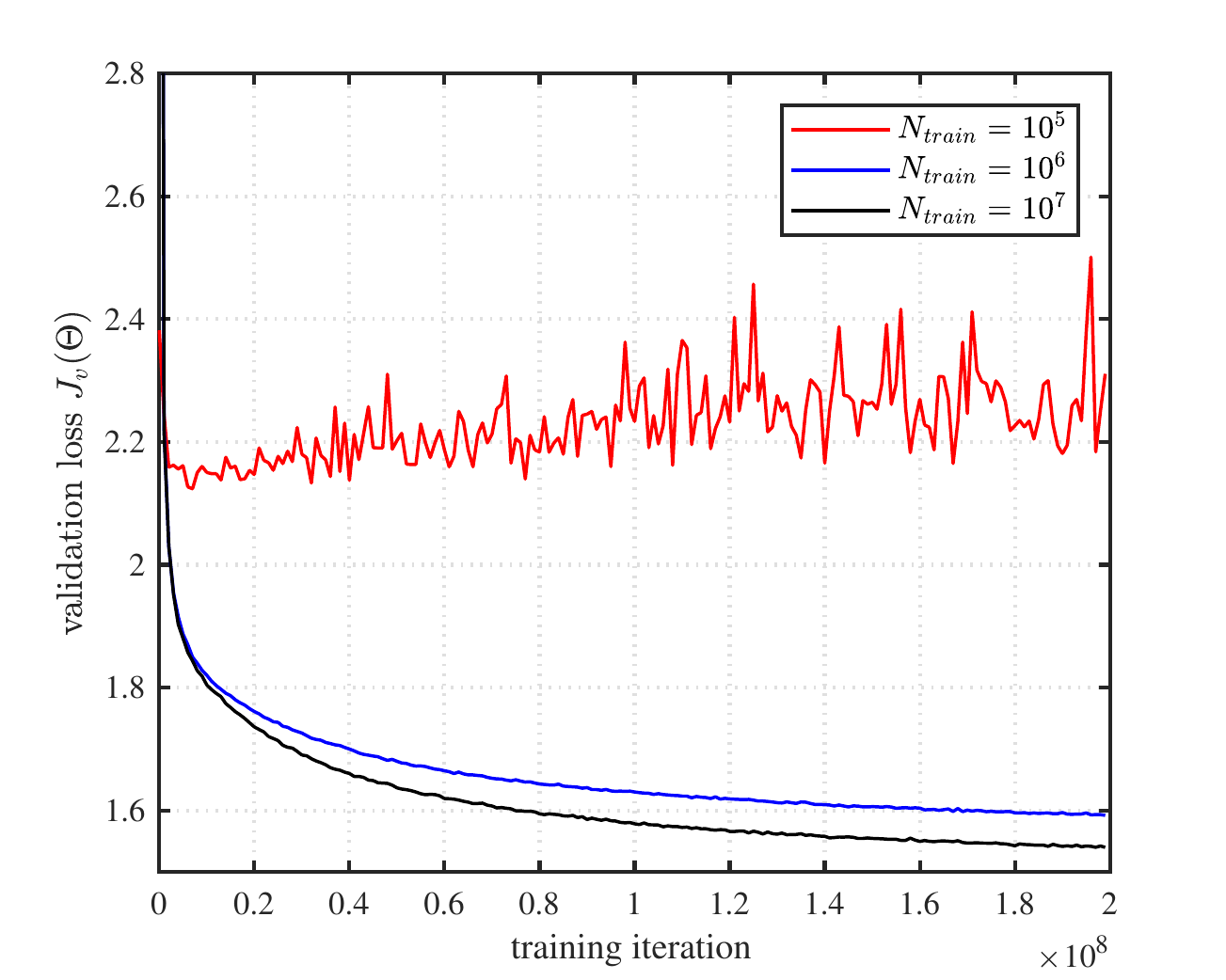}
\caption{Validation loss $J_{v}(\Theta)$ for various number of training samples $N_{train}$ ($k=4$ and $m=70$). }
\label{fig:sim_validation_loss}
\end{figure}
In this section, we go over two major issues when applying the D-AUD scheme in the practical scenarios.
We first discuss the training data collection issue.
This issue is crucial since the uplink traffics are usually unpredictable and sporadic so that it takes quite a bit of time and effort to collect the training data.
We next discuss a sparsity estimation issue.
In order to perform the symbol detection and decoding, the basestation should know the sparsity (number of active devices) in advance.

\subsection{Training Data Collection}
In order to learn the optimal mapping function $g^{\ast}$ between the received signal and support, sufficient amount of training data is required.
In Fig.~\ref{fig:sim_validation_loss}, we plot the validation loss $J_{v}(\Theta)$ as a function of the training iteration for various sizes of training dataset.
We see that when the number of training dataset is not enough, the deep neural network does not converge, causing a failure in the D-AUD training.
In acquiring the dataset, it would be natural to use the \textit{real} received signals, but it requires too many training data transmissions.
For example, when collecting one million received signals in LTE systems, it will take around 30 minutes (1 ms subframe consisting of 14 symbols).
This time will further increase in proportion to the number of ensemble networks.
Therefore, this type of data collection is by no means practical in terms of energy consumption, latency, and resource utilization efficiency.

In order to reduce the overhead associated with the training data transmission, we use  synthetically generated signals as the training data set at basestation.
One might concern that the synthetically generated signal is different from the actual transmitted signal since the channel depends heavily on the environmental factors such as frequency band, mobility, and geometric objects.
Fortunately, we can circumvent this issue since the AUD process is essentially the same as the support identification and all channel components are contained in an input sparse vector $\mathbf{x}$, not the system matrix $\mathbf{\Phi}$ (see \eqref{eq:system_model}).
Thus, the D-AUD scheme only needs to learn the codebook matrix $\mathbf{\Phi}$ (which is known a priori), not the individual channel states, which will ease the training process significantly.
Indeed, what we need to do in the training phase is to artificially generate the received vector in \eqref{eq:system_model}.
In doing so, time and effort to collect huge training data can be saved and at the same time the training process can be done offline.

Since the training operation of D-AUD is performed offline using the synthetically generated data, we train multiple D-AUD networks for various settings in terms of the number of total users and the number of active users.
From this process, we can obtain the internal parameters (e.g., weight and bias) for each scenario.
When applying the D-AUD to the actual transmission, we thus use the pre-trained network corresponding to the system environment.
Even though the system environment will vary, as long as we have a trained model matching to the environment, re-training process is unnecessary.
For example, we train two D-AUD schemes for two different number of total users ($N=50,$ and 100). In the test phase, when $N$ changes from 30 to 80, what we need to do is to change the trained model for $N=50$ to that for $N=100$.

\subsection{Sparsity Estimation}
In the grant-free transmission, devices can transmit the data without the granting process so that the basestation needs to be aware of the sparsity to perform the AUD.
Since the sparsity is used as the number of iterations in many sparse recovery algorithms, incorrect sparsity leads to either miss detection (early termination) or false alarm (late termination).
In the former case, some of active devices cannot be identified while inactive devices can be chosen as active devices for the latter case.
Therefore, the sparsity estimation error degrades the support identification quality substantially\footnote{As a sparsity estimation strategy, the residual-based stopping criterion is widely used [24]. In this scheme, basically, an algorithm is terminated when the residual power $\lVert \mathbf{r} \rVert_{2}$ is smaller than the pre-specified threshold $\epsilon$ (i.e., $\lVert \mathbf{r} \rVert_{2} < \epsilon$) and the iteration number at the termination point is set to the sparsity level. However, since the residual magnitude decreases monotonically and the rate of decay depends on the system parameters, it might not be easy to figure out an accurate terminating point.}.


\begin{algorithm}[t]
\renewcommand{\thealgorithm}{}
\floatname{algorithm}{}
	\caption{\textbf{Algorithm 1.} Sparsity estimation in the proposed D-AUD scheme}
	\begin{algorithmic}[1]
	\INPUT 
	the received signal $\hat{\mathbf{y}}\in\mathbb{R}^{2m\times 1}$, the trained threshold $\tau\in\mathbb{R}$, the maximal sparsity $K\in\mathbb{R}$
	\OUTPUT the estimated sparsity $\hat{k}$, the estimated support $\hat{\Omega}$
	\INITIAL $l=0$, $\Gamma = \{1,\cdots,N\}$
	\WHILE {$l \leq K$ and $l \neq \vert \Gamma \vert$}
	\STATE $l=l+1$
	\STATE Obtain $\hat{\mathbf{p}}^{(l)}$ by passing $\hat{\mathbf{y}}$ into the D-AUD network trained for sparsity level $l$
	\STATE $\hat{p}_{\max}^{(l)}=\underset{i}{\max} \ \hat{p}_{i}^{(l)}$
   	\STATE $\Gamma = \left\{ i\in\{1,\cdots,N\} \ \Big\vert \ \frac{\hat{p}_{i}^{(l)}}{\hat{p}_{\max}^{(l)}}\geq \tau \right\}$
   	\ENDWHILE
   	\STATE $\hat{k} = l$
   	\STATE $\hat{\Omega} = \Gamma$
   	\RETURN $\hat{k}$, $\hat{\Omega}$
	\end{algorithmic}
	\label{alg:decoding}
\end{algorithm}

In the proposed D-AUD scheme, instead of using an iterative support identification, $k$ support elements are chosen from the softmax output (see \eqref{eq:final_support}).
Thus, in contrast to the conventional sparse recovery algorithms, a separate sparsity estimation process is unnecessary.
One simple option to choose the support is to take the indices of the softmax output values being larger than the threshold $\tau$.
Benefit of this approach is that $\tau$ can be readily chosen in the training phase since the support $\Omega$ and sparsity $k$ of training data are already available.
When determining $\tau$, we use both softmax output $\hat{\mathbf{p}}$ and sparsity $k$ of training data.
Specifically, in the training phase, we obtain the softmax values $\hat{p}_{\omega_{1}},\cdots,\hat{p}_{\omega_{k}}$ for $\omega_{i}\in\Omega$.
Note that these values would be close to $\frac{1}{k}$ since the D-AUD is trained to generate the true probability $p_{\omega_{i}}=\frac{1}{k}$. 
In order to remove the effect of $k$, meaning that $\tau$ is set to be independent of $k$, we scale $\hat{p}_{\omega_{1}},\cdots,\hat{p}_{\omega_{k}}$ by $k$ and then set the minimum value to $\tau$ (i.e., $\tau=\underset{i}{\min}\ k\hat{p}_{\omega_{i}}$).
In doing so, in the test phase, we can identify the support without the knowledge of the sparsity $k$.
To be specific, by using multiple D-AUD networks for $K$ sparsity levels, we obtain $K$ softmax output vectors $\hat{\mathbf{p}}^{(l)}=[\hat{p}^{(l)}_{1},\cdots,\hat{p}^{(l)}_{N}]$ for $l=1,\cdots, K$.
Then, we take indices satisfying $\frac{\hat{p}^{(l)}_{i}}{\hat{p}^{(l)}_{\text{max}}} \geq \tau$ for $i=1,\cdots,N$ where $\hat{p}^{(l)}_{\text{max}}=\underset{i}{\max}\ \hat{p}^{(l)}_{i}$ is the maximal value of $\hat{\mathbf{p}}^{(l)}$.
If the number of the chosen indices is the same as $l$, then we find out the estimated sparsity $\hat{k}=l$ and the estimated support $\hat{\Omega}$.
The proposed sparsity estimation in the D-AUD scheme is summarized in Algorithm 1.

\section{Simulations and Discussions}

\subsection{Simulation Setup}
In this section, we investigate the performance of the proposed D-AUD scheme.
Our simulation setup is based on the grant-free NOMA transmission in the orthogonal frequency division multiplexing (OFDM) systems specified in the 3GPP NR Rel.16~\cite{3GPPNOMA}.
Specifically, we use 100 users ($N=100$) and 70 subcarriers ($m=70$) in each transmission so that the overloading factor is 143\%.
As a channel model, the pathloss component $\gamma_{i}$ between the $i$-th device and the basestation is modeled as $\gamma_{i}=128.1+37.6\log_{10}(d_{i})$ [dB scale] where $d_{i}$ is the distance (in km) between the $i$-th device and the base station~\cite{channel_model} and independent Rayleigh fading coefficient is used for each device~\cite{channel_model_2}.
The noise spectral density and transmission bandwidth are set to -170 dBm/Hz and 1 MHz, respectively.
For comparison, we examine the performance of the conventional LS-BOMP~\cite{CStrick}, MMSE-BOMP~\cite{MMSE_BOMP_s}, and approximate message passing (AMP) algorithm~\cite{AMP_MIMO}.
When generating nonzero values in the LDS codebook, we use an i.i.i. Gaussian random variable.
Length of the LDS codeword $S$ is set to 10 ($S=10$).

\begin{figure}[t]
	\centering
    \includegraphics[width=.6\columnwidth]{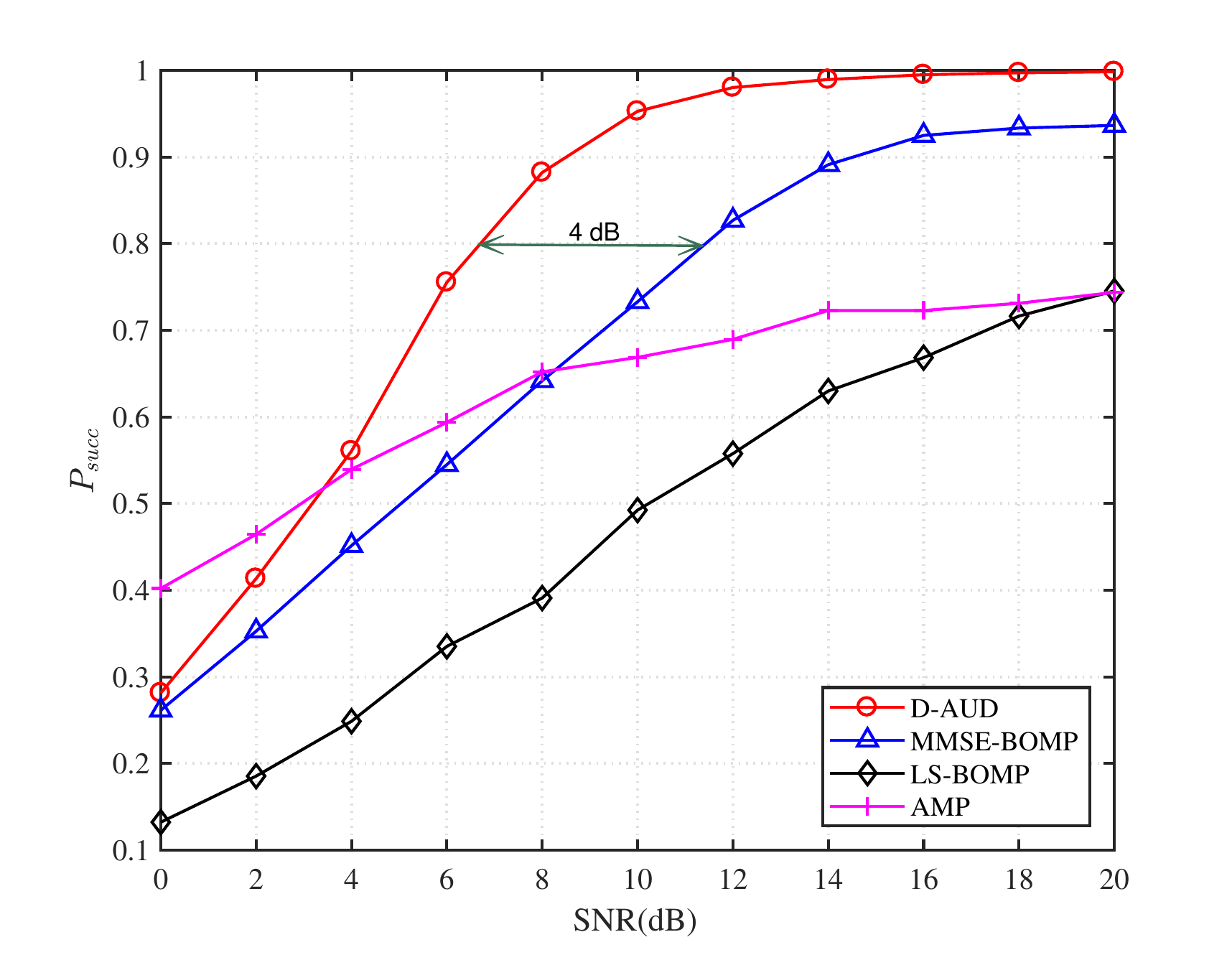}
    \caption{$P_{succ}$ as a function of SNR ($N=100, k=4, N_{d}=7, m=70$).}
    \label{fig:sim_1}
\end{figure}

In order to guarantee the model stability of the D-AUD scheme, we use $K$-fold cross validation in the training phase.
In the $K$-fold cross validation, total samples are randomly partitioned into $K$ equal-sized sets.
Among $K$ partitioned sets, a single set is used for the model testing, and the remaining $K-1$ sets are used for the D-AUD training.
Then, this process is repeated $K-1$ times for the remaining sets.
In our simulations, we generate ${10^7}$ samples and set $K=10$.
When selecting the hyperparameters, we use the cross-validation technique (see Fig. 12).
In our simulations, we use an Adam optimizer, well-known optimization tool to guarantee the robustness of learning process~\cite{OP}.
As an activation function in hidden layers, we used ReLU function (i.e., $f_{ReLU}(x)=\max(0,x)$).
Also, we set $L=6$ (the number of hidden layers), $\alpha=1000$ (the width of hidden layer), $P_{drop}=0.1$ (dropout probability), $\eta = 5\times 10^{-4}$ (learning rate), and $E=3$ (the number of ensemble networks).
As a performance metric, we use the success probability $P_{succ}$ which corresponds to the percentage of the detected users among all active users.

\subsection{Simulation Results}

\begin{figure}[t]
	\centering
    \includegraphics[width=.6\columnwidth]{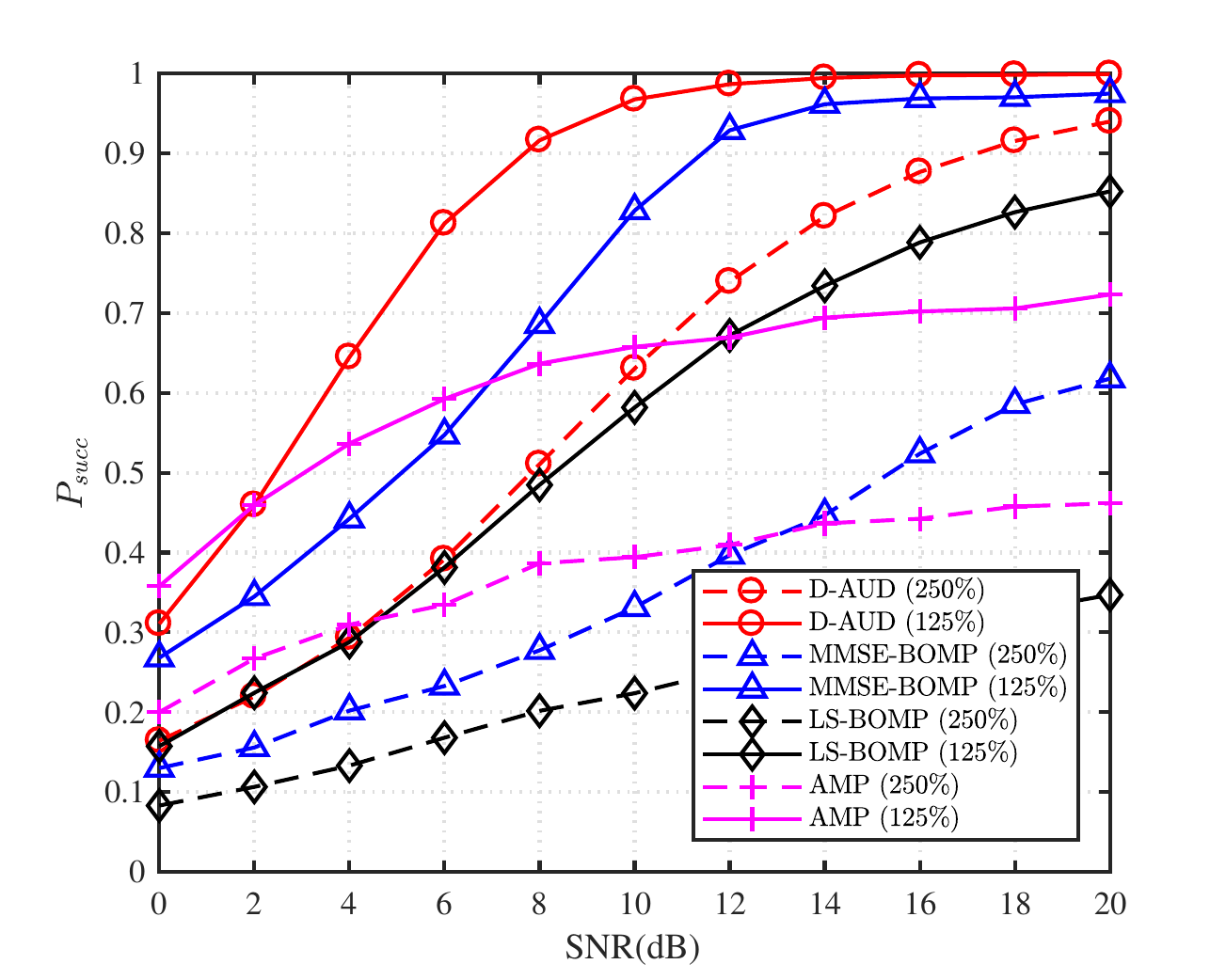}
    \caption{$P_{succ}$ as a function of SNR with various overloading factor ($N=100, k=4)$.}
    \label{fig:sim_2}
\end{figure}

In Fig.~\ref{fig:sim_1}, we evaluate $P_{succ}$ of the proposed D-AUD scheme and competing AUD schemes as a function of SNR.
We observe that D-AUD outperforms the conventional schemes for all SNR regime.
Since D-AUD learns the mapping between the received signal $\tilde{\mathbf{y}}$ and the support $\Omega$, an estimate of support $\hat{\Omega}$ can be determined only by the input data $\tilde{\mathbf{y}}$.
This means that the whole AUD process can be handled by a simple end-to-end mapping in D-AUD.
For example, we observe that D-AUD achieves around 6 dB gain over the MMSE-BOMP at $P_{succ} = 0.9$.

\begin{figure}[t]
	\centering
    \includegraphics[width=.6\columnwidth]{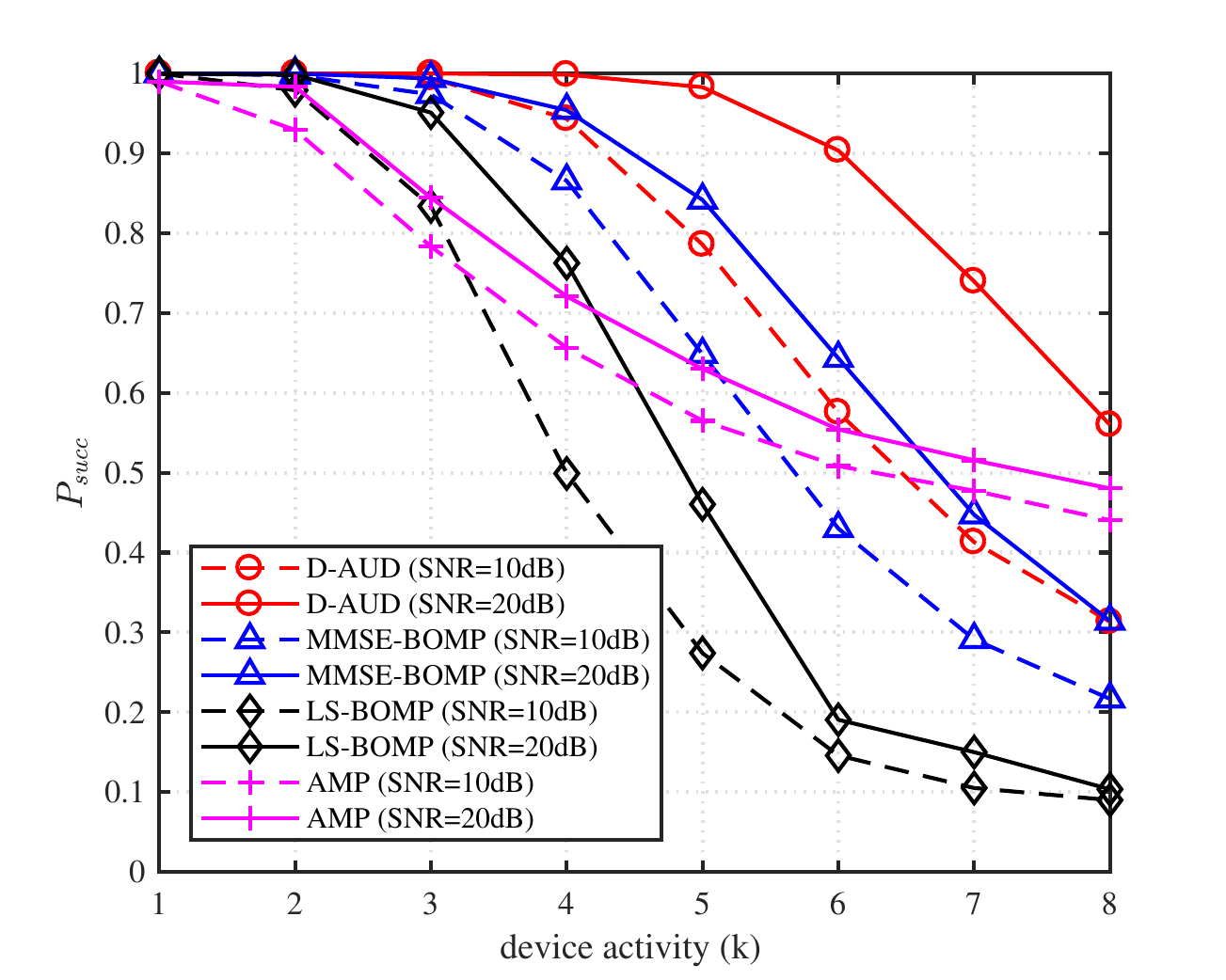}
    \caption{$P_{succ}$ as a function of $k$ with 2 different SNR ($N=100, N_{d}=7, m=70$).}
    \label{fig:sim_3}
\end{figure}

In Fig.~\ref{fig:sim_2}, we investigate $P_{succ}$ for various overloading factors.
We can clearly see that D-AUD outperforms the conventional AUD approaches by a large margin.
For example, in case of 125$\%$ overloading, D-AUD achieves around 4 dB gain over the MMSE-BOMP at $P_{succ}=0.9$.
We also observe that the AUD performance of D-AUD is robust to the overloading factor due to the decoupling of the correlated activation patterns (see Section III).
For instance, in case of 250$\%$ overloading, D-AUD achieves $P_{succ}=0.9$ at SNR $=17.5$ dB.
Since there is no such mechanism for the conventional sparse recovery algorithms, performance of conventional schemes is not appealing when the overloading factor is high.

In Fig.~\ref{fig:sim_3}, we plot the $P_{succ}$ as a function of the number of active 
devices $k$.
We observe that D-AUD outperforms conventional schemes across the board.
For example, when the number of active devices is 6 (i.e., $k=6$) and SNR $= 20$ dB, $P_{succ}$ of D-AUD is 0.9 while those of the MMSE-BOMP and AMP are 0.65 and 0.55, respectively.
Also, we can see that the D-AUD maintains its robustness even when $k$ increases.
When $k$ increases, the mutual correlation associated with the active devices becomes large, causing a severe degradation of the AUD performance (see Section II).
Since the DNN already learned the correlation feature from the training dataset, D-AUD can better discriminate the correlated supports in the test phase.
For example, when $k$ increases from 3 to 6, $P_{succ}$ of the D-AUD decreases marginally from 0.99 to 0.91.
However, $P_{succ}$ of the MMSE-BOMP and LS-BOMP decrease sharply from 0.98 to 0.65 and from 0.95 to 0.19, respectively.

\begin{figure}[t]
\centering
\subfloat[]{\includegraphics[width=.33\linewidth]{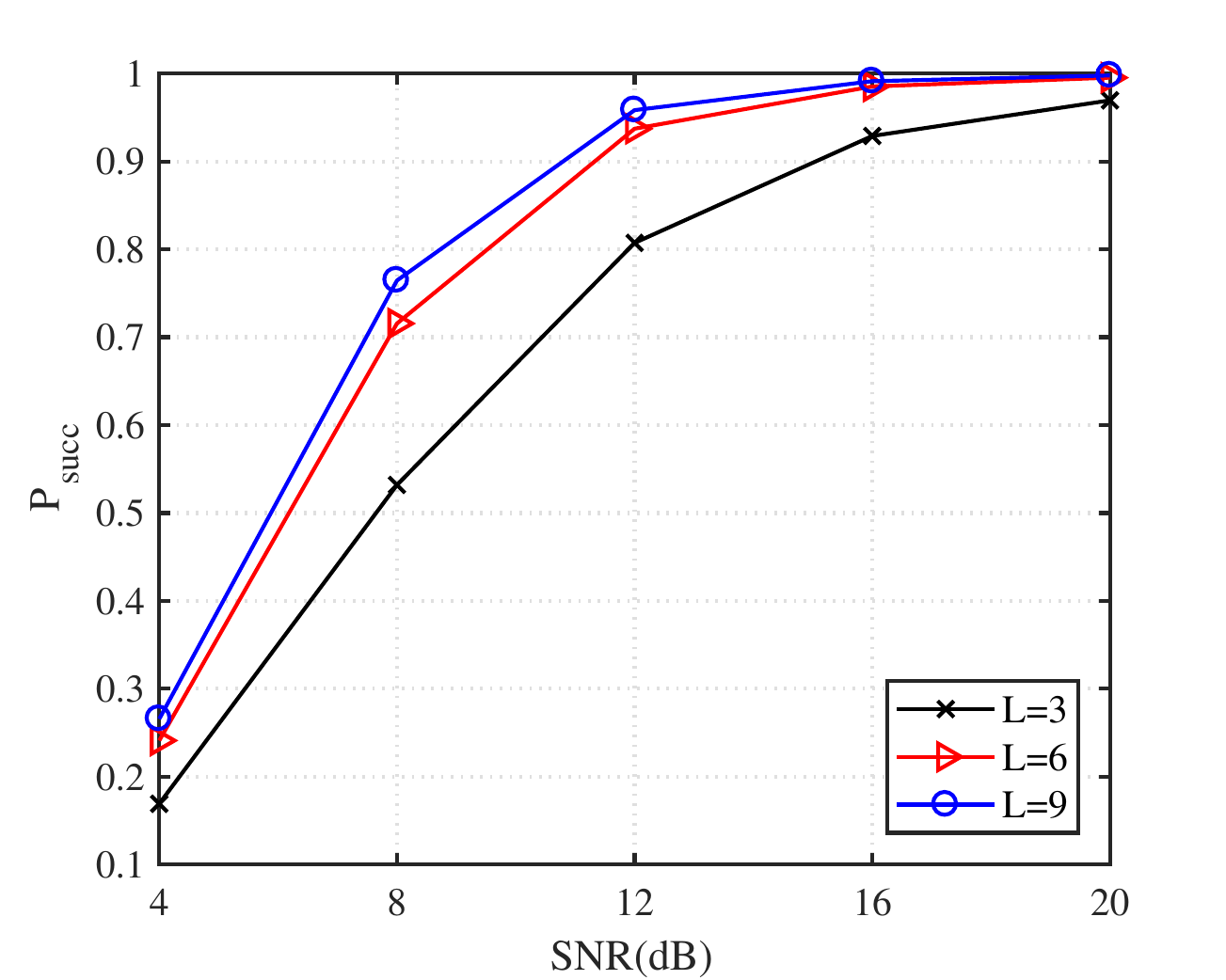}}
\subfloat[]{\includegraphics[width=.33\linewidth]{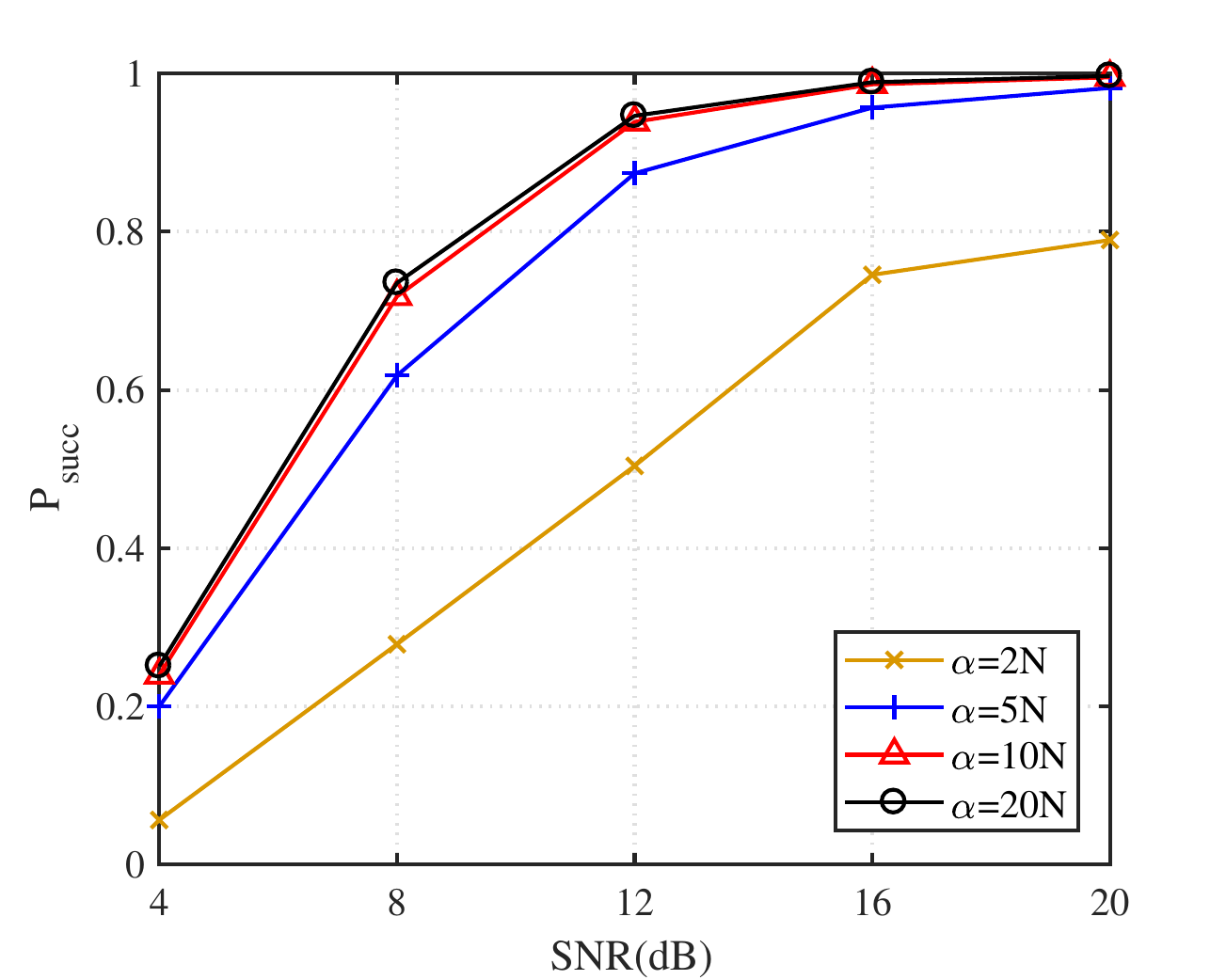}}
\subfloat[]{\includegraphics[width=.33\linewidth]{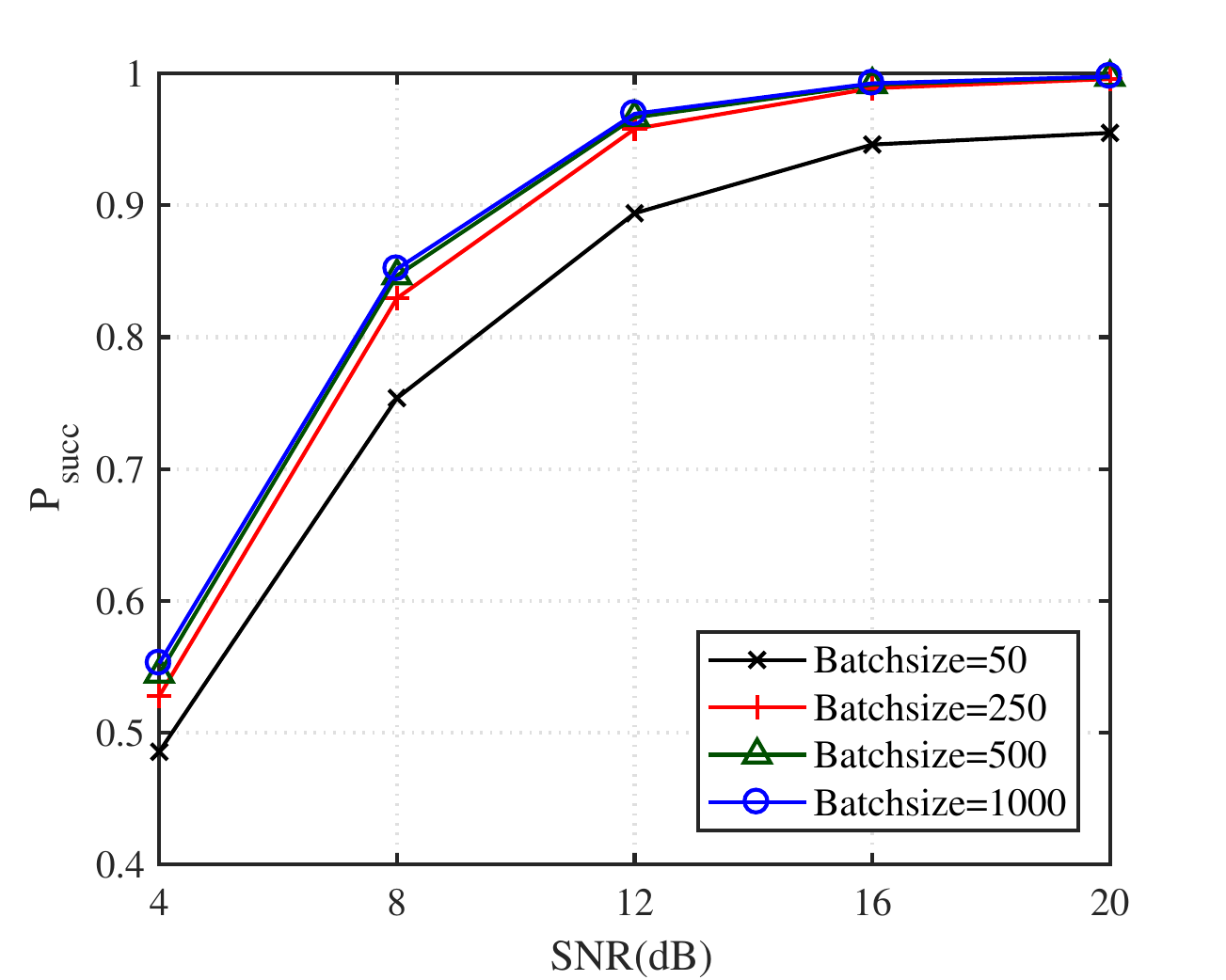}}
\hfill
\subfloat[]{\includegraphics[width=.33\linewidth]{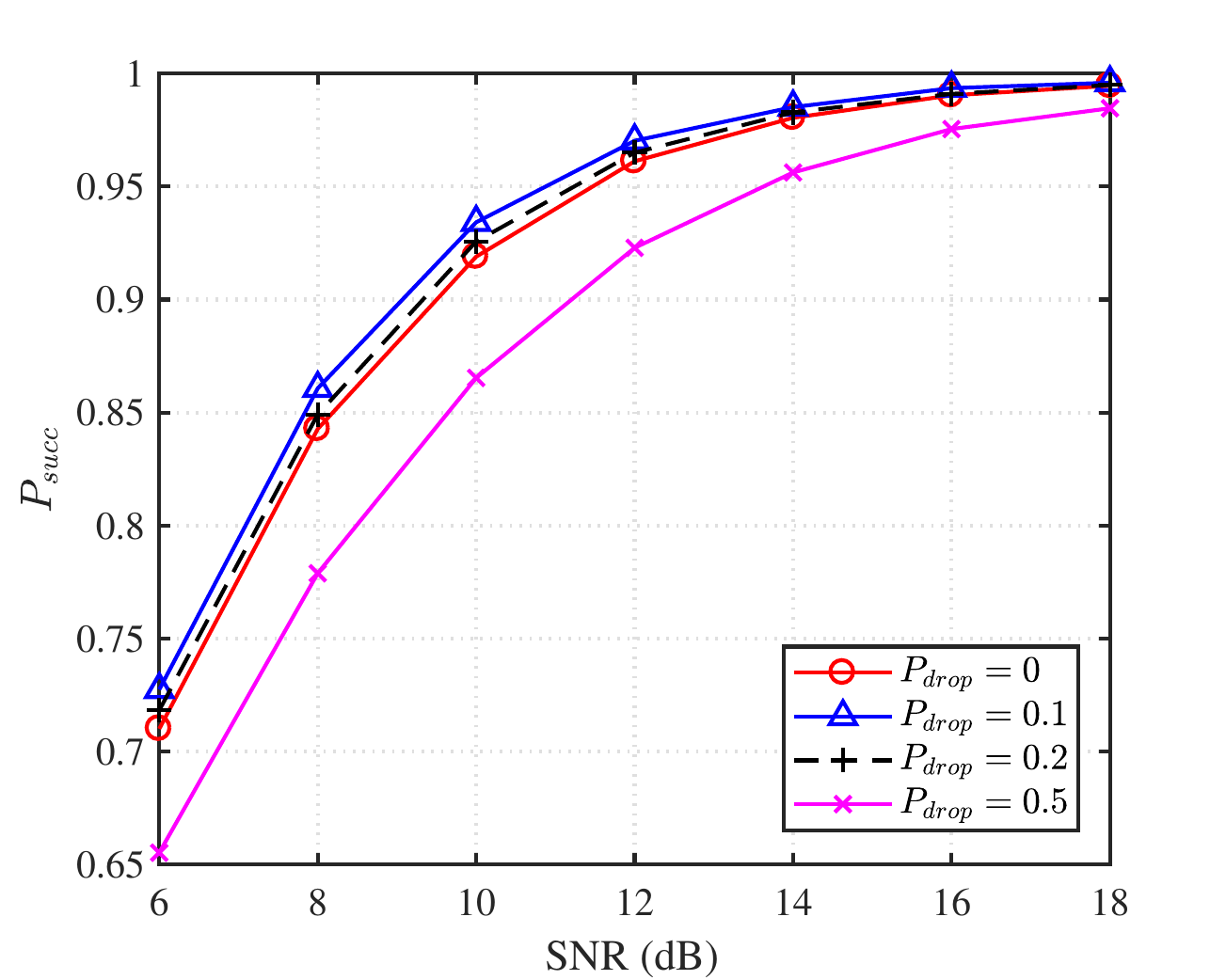}}
\subfloat[]{\includegraphics[width=.33\linewidth]{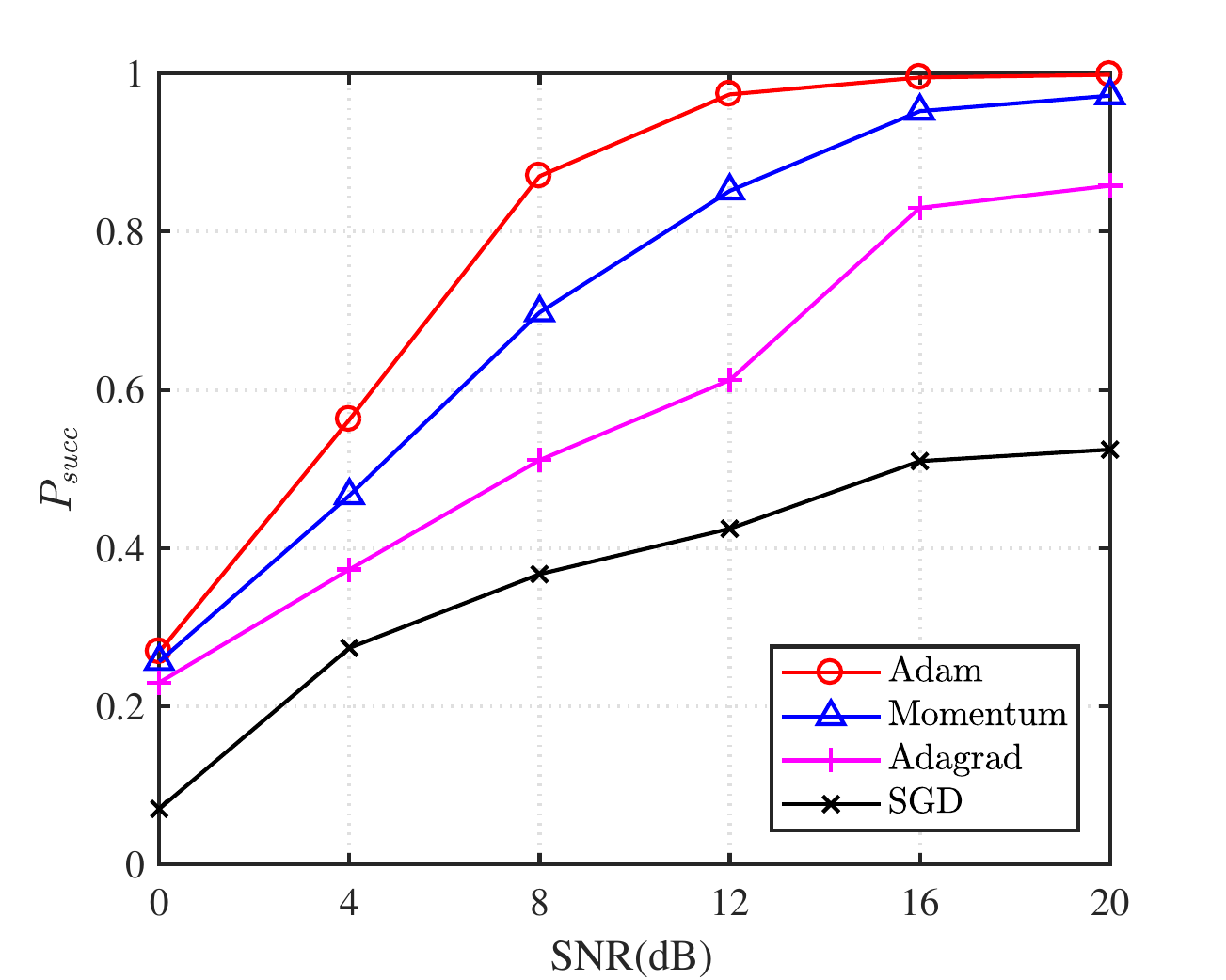}}
\subfloat[]{\includegraphics[width=.33\linewidth]{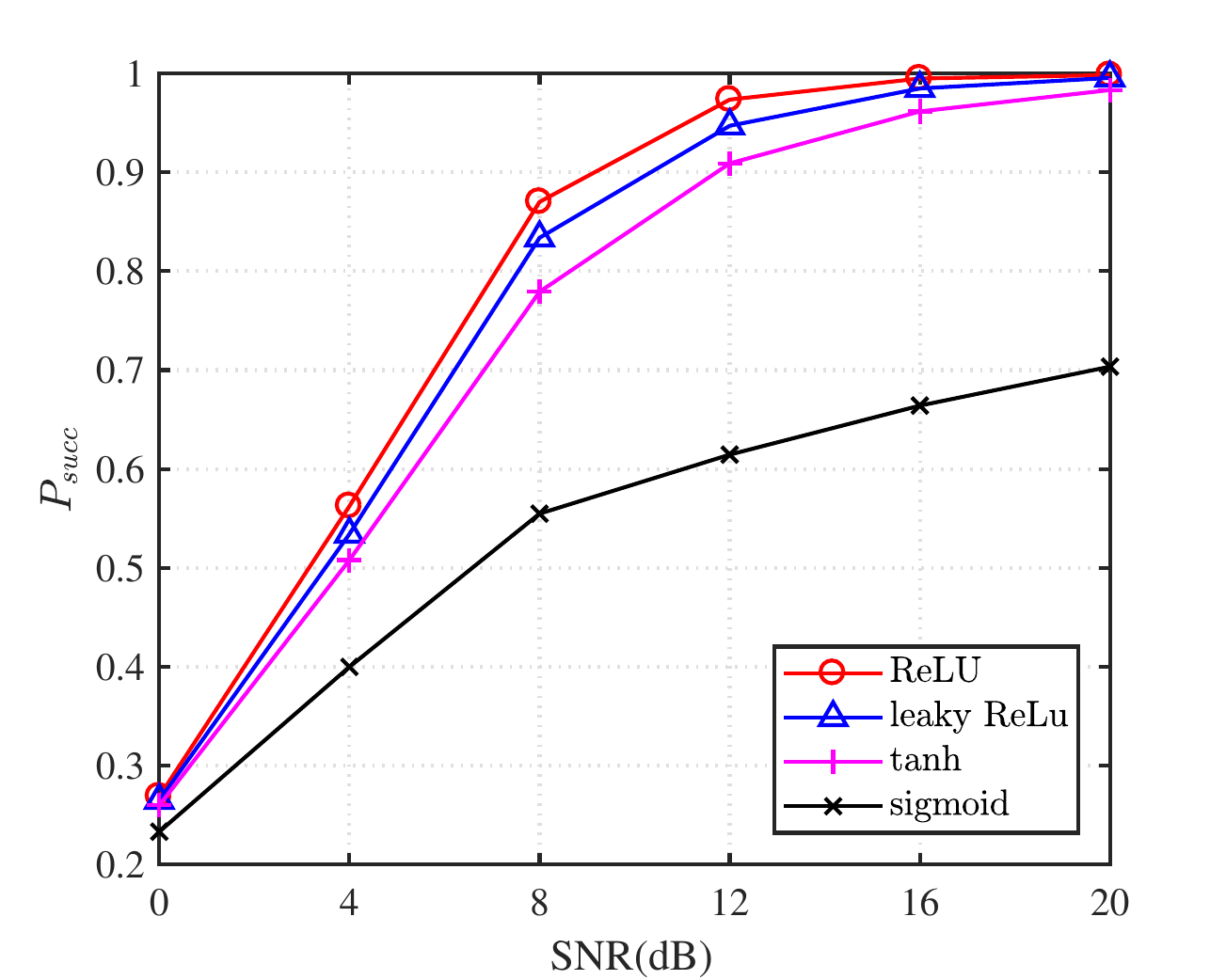}}
\caption{An example of hyperparameter tuning process: (a) depth of hidden layers, (b) width of hidden layers, (c) batch size, (d) dropout probability, (e) optimizer, and (f) activation function}
\label{fig:hyperparameter}
\end{figure}

In Fig. 12, we evaluate $P_{succ}$ for various hyperparameters such as depth and with of hidden layers, batch size, dropout probability, activation function, and optimizer.
From these results, we can observe the effect of each hyperparameter on the AUD performance.
For example, if the width of the hidden layer is small (e.g., $\alpha = 2N$ case), we expect that the performance of D-AUD will degrade considerably.
Whereas, if the width is larger than $10N$, the D-AUD performance will not be improved further.
From this offline tuning process, we can obtain the best hyperparameters of D-AUD.

\begin{figure}[t]
	\centering
    \includegraphics[width=.6\columnwidth]{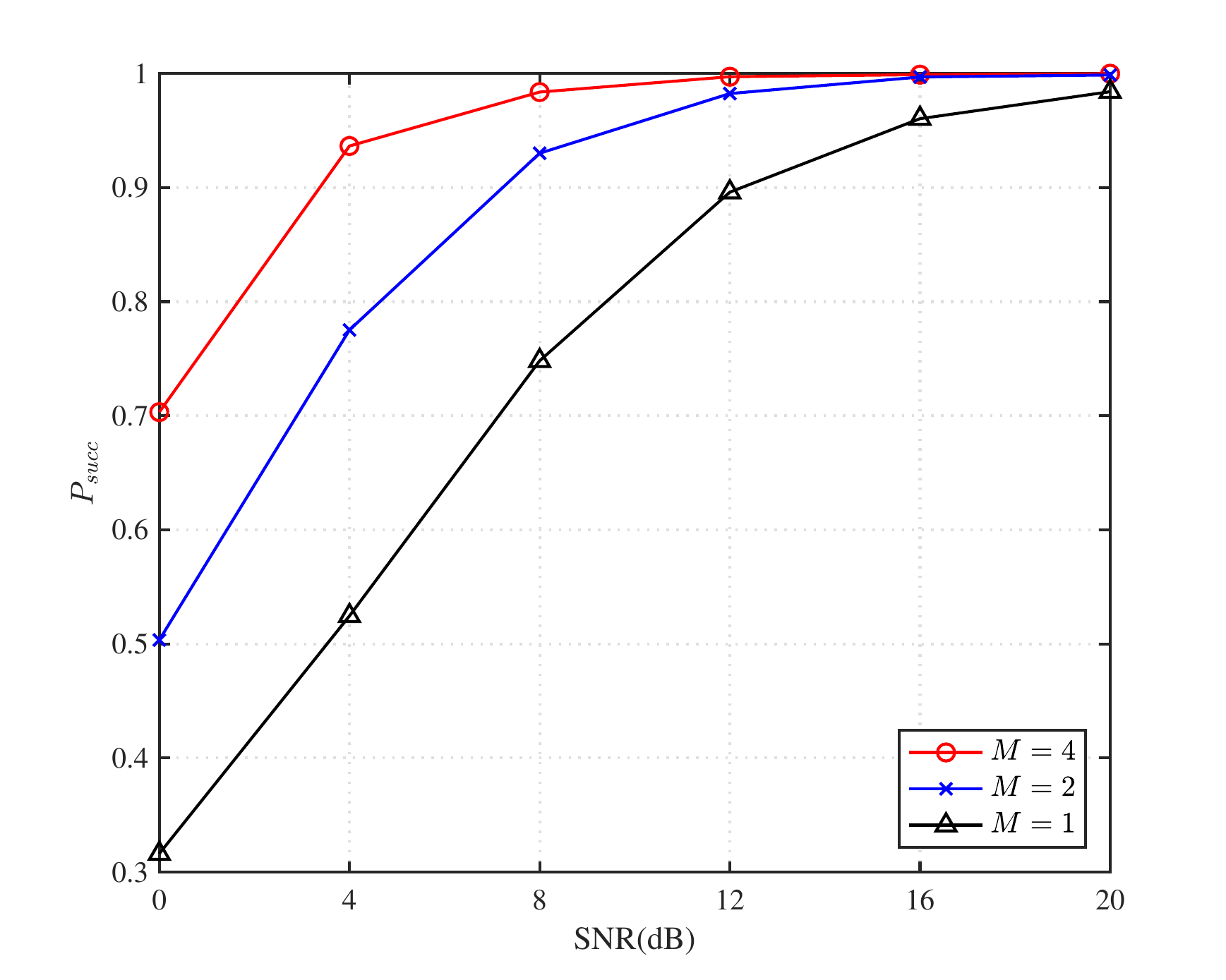}
    \caption{$P_{succ}$ as a function of SNR in the multi-antenna scenario ($N=100, k=4$).}
    \label{fig:sim_4}
\end{figure}

Finally, in order to test the performance of D-AUD scheme in multiple-antenna scenarios\footnote{When using $M$ antenna at the base station, the input of the D-AUD scheme will become multiple measurement vectors $\tilde{\mathbf{y}}_{(1)},\cdots,\tilde{\mathbf{y}}_{(M)}$, not a single measurement vector $\tilde{\mathbf{y}}$. Accordingly, the AUD problem, originally modeled as single measurement vector (SMV) problem, will be also converted to the multiple measurement vector (MMV) problem. From the equation (5), we can obtain the MMV model $\tilde{\mathbf{Y}}=\boldsymbol{\Phi}\mathbf{X}$ where $\tilde{\mathbf{Y}} = [\tilde{\mathbf{y}}_{(1)} \ \cdots \ \tilde{\mathbf{y}}_{(M)}]$ and $\mathbf{X} = [\mathbf{x}_{(1)} \ \cdots \ \mathbf{x}_{(M)}]$. Since the supports of $\mathbf{x}_{(1)}, \cdots, \mathbf{x}_{(M)}$ are common, we can exploit the correlation among them in the recovery process.}, we consider the three distinct cases (i.e., number of received antennas is $M=1, 2,$ and 4).
As shown in Fig. 13, we observe that the performance of D-AUD improves with $M$.
For example, when $M = 4$, we observe 8.4 dB gain over the single received antenna scenario ($M=1$) at $P_{succ}=0.9$.
Since the active users are detected blindly (without the channel information), the base station cannot achieve the gain proportional to the number of antennas.
Nevertheless, the multi-antenna gain proportional to the number of antennas (around 2.1 dB gain per antenna) can be achieved.

\section{Conclusion}
In this paper, we proposed a novel AUD scheme called D-AUD based on the DNN for the mMTC uplink scenario.
Our work is motivated by the observation that CS-based AUD cannot support the massive number of devices and high device activity scenario in the grant-free NOMA systems.
By applying the training data to the properly designed DNN, the proposed D-AUD scheme learns the nonlinear mapping between the received signal and support.
As long as we train the deeply stacked hidden layers with a proper loss function, we can detect active devices in the test phase.
We demonstrated from numerical evaluations that the proposed D-AUD scheme is very effective in the highly-overloaded mMTC scenarios.
In this paper, we restricted our attention to the AUD but we believe that there are many interesting applications of the proposed approaches such as DoA estimation, mmWave channel estimation, and MIMO detection.

\begin{appendices}
\section{Proof of the computational complexities in Table I}
In this appendix, we analyze the computational complexities of LS-BOMP and MMSE-BOMP in Table I.
We first analyze the complexity of LS-BOMP.
In the $j$-th iteration of LS-BOMP, a submatrix $\mathbf{\Phi}_{l}$ of $\mathbf{\Phi}$ having the maximum correlation between the residual vector $\mathbf{r}^{j-1}$ is chosen (see \eqref{eq:bomp_iteration}).
The corresponding complexity $\mathcal{C}_{\mathrm{I}}$ is
\begin{align}
\mathcal{C}_{\mathrm{I}} = \sum_{j=1}^{k} {\{ (2m-1)mN + (mN-1) \} } = 2km^2 N-k.
\label{eq:lsbomp_identification}
\end{align}
After identifying a support element, a signal vector $\mathbf{x}^{j}$ is estimated using the LS estimator (i.e., $\mathbf{x}^{j} = \left( \mathbf{\Phi}_{\Omega_{j}}^{H}\mathbf{\Phi}_{\Omega_{j}} \right)^{-1}\mathbf{\Phi}_{\Omega_{j}}^{H} {\mathbf{y}}$).
Using the Cholesky decomposition~\cite{Cholesky}, the resulting computational complexity $\mathcal{C}_{\mathrm{LS}}$ is approximated as
\begin{align}
\mathcal{C}_{\mathrm{LS}}
&\approx \sum_{j=1}^{k} {(m+ \frac{jm}{3})j^2 m^2} \\
&= \frac{k^4+6k^3+7k^2+2k}{12} m^3.
\label{eq:lsbomp_estimation}
\end{align}
Finally, the residual vector $\mathbf{r}^{j-1}$ is updated as $\mathbf{r}^{j}= \mathbf{y}-\mathbf{\Phi}_{{\Omega}^{j}} \hat{\mathbf{x}}^{j}$.
The corresponding complexity $\mathcal{C}_{\mathrm{U}}$ is
\begin{align}
\mathcal{C}_{\mathrm{U}} = \sum_{j=1}^{k} {\{ (2jm-1)m+m \}} = k(k+1)m^2.
\label{eq:lsbomp_update}
\end{align}
From ~\eqref{eq:lsbomp_identification} to ~\eqref{eq:lsbomp_update}, the complexity $\mathcal{C}_{\mathrm{LS-BOMP}}$ of LS-BOMP is
\begin{align}
\mathcal{C}_{\mathrm{LS-BOMP}}
&=
\mathcal{C}_{\mathrm{I}} + \mathcal{C}_{\mathrm{LS}} + \mathcal{C}_{\mathrm{U}} \\
&=
2km^2 N-k + \frac{k^4+6k^3+7k^2+2k}{12} m^3 + k(k+1)m^2.
\end{align}

Now, we analyze the complexity of MMSE-BOMP.
Since the support identification and residual update of MMSE-BOMP are the same as those of LS-BOMP, the corresponding complexities ($\mathcal{C}_{\mathrm{I}}$ and $\mathcal{C}_{\mathrm{U}}$) are also the same as LS-BOMP.
When estimating the signal values, the MMSE estimator is used (i.e., $\mathbf{x}^{j} = \mathbf{\Phi}_{\Omega_{j}}^{H}  \left( \mathbf{\Phi}_{\Omega_{j}} \mathbf{\Phi}_{\Omega_{j}}^{H} + \frac{\sigma_{n}^{2}}{\sigma_{x}^{2}} \mathbf{I} \right)^{-1} \mathbf{y}$).
By approximating the complexity of the matrix inversion operation~\cite{Inversion}, the resulting complexity $\mathcal{C}_{\mathrm{MMSE}}$ is
\begin{align}
\mathcal{C}_{\mathrm{MMSE}}
&\approx \sum_{j=1}^{k} {\left\{2m + j \left( \frac{14}{3}m^3+m^2-m \right) \right\}} \\
&= 2km + \frac{k(k+1)}{2} \left( \frac{14}{3}m^3+m^2-m \right).
\label{eq:mmsebomp_estimation}
\end{align}
The resulting complexity $\mathcal{C}_{\mathrm{MMSE-BOMP}}$ of MMSE-BOMP is
\begin{align}
\mathcal{C}_{\mathrm{MMSE-BOMP}}
&=
\mathcal{C}_{\mathrm{I}} + \mathcal{C}_{\mathrm{MMSE}} + \mathcal{C}_{\mathrm{U}} \\
&=
2km^2 N-k + 2km + \frac{k(k+1)}{2} \left( \frac{14}{3}m^3+m^2-m \right) + k(k+1)m^2
\end{align}
\end{appendices}

\bibliographystyle{ieeetr}
\bibliography{Reference}

\end{document}